\documentclass[11pt,A4]{article}

\usepackage[english]{babel}
\usepackage[utf8]{inputenc}
\usepackage{amssymb,amsfonts,amsmath}
\usepackage{graphicx}
\usepackage{authblk}
\usepackage[framed,numbered,autolinebreaks,useliterate]{mcode}

\usepackage{newfloat}
\DeclareFloatingEnvironment[name={Supplementary Figure}]{suppfigure}
\DeclareFloatingEnvironment[name={Supplementary Table}]{supptab}

\DeclareMathOperator*{\sgn}{sgn}

\begin{document}
\title{Dynamic similarity promotes interpersonal coordination in joint action}

\author[1]{Piotr S\l{}owi\'{n}ski}
\author[2]{Chao Zhai}
\author[2]{Francesco Alderisio}
\author[3]{Robin Salesse}
\author[3]{Mathieu Gueugnon}
\author[3]{Ludovic Marin}
\author[3,4]{Benoit G.~Bardy}
\author[2,5,+]{Mario di~Bernardo}
\author[1,+,*]{Krasimira Tsaneva-Atanasova}

\affil[1]{Department of Mathematics, College of Engineering, Mathematics and Physical Sciences, University of Exeter, EX4 4QF, United Kingdom}
\affil[2]{Department of Engineering Mathematics, University of Bristol, Merchant Venturers' Building, BS8 1UB, United Kingdom}
\affil[3]{EuroMov, Montpellier University, 700 Avenue du Pic Saint-Loup, 34090 Montpellier, France.}
\affil[4]{Institut Universitaire de France, Paris, France}
\affil[5]{Department of Electrical Engineering and Information Technology, University of Naples Federico II, 80125 Naples, Italy}
\affil[+]{These authors contributed equally to this work} 
\affil[*]{Corresponding author's email: K.Tsaneva-Atanasova@exeter.ac.uk}
\date{}

\maketitle

\begin{abstract}
Human movement has been studied for decades and dynamic laws of motion that are common to all humans have been derived. Yet, every individual moves differently from everyone else (faster/slower, harder/smoother etc). We propose here an index of such variability, namely an individual motor signature (IMS) able to capture the subtle differences in the way each of us moves. 
We show that the IMS of a person is time-invariant and that it significantly differs from those of other individuals. This allows us to quantify the dynamic similarity, a measure of rapport between dynamics of different individuals' movements, and demonstrate that it facilitates coordination during interaction. 
We use our measure to confirm a key prediction of the theory of similarity that coordination between two individuals performing a joint-action task is higher if their motions share similar dynamic features. Furthermore, we use a virtual avatar driven by an interactive cognitive architecture based on feedback control theory to explore the effects of different kinematic features of the avatar motion on the coordination with human players.

\end{abstract}

\section{Introduction}

Humans often need to perform joint tasks and coordinate their movement \cite{Schmidt2008}. Their motion can be studied and classified by means of some common generalised movement laws \cite{Hogan1987,Viviani1995,Hogan2007} that define a ``human-like'' way of movement \cite{Kilner2003,Kilner2007}. However, every individual moves following a specific personal style characterised by unique kinematic features. An open problem is to find methods that capture such features and identify different individuals from the way they move. This is key to show that individuals moving in similar ways exhibit higher levels of synchronisation when performing joint tasks.

Specifically, studies of interpersonal interaction show that people prefer to interact with others who are similar to themselves \cite{Folkes1982,Marsh2009}. Moreover, it has been shown that social movement coordination between interacting people could be used to assess their mutual rapport \cite{Lakens2011,Oullier2008,Schmidt2014}. These observations have led to the development of a theory of similarity which predicts that the level of synchronisation in joint actions is enhanced if the participants are similar in terms of morphology and movement dynamics and are willing to match their behaviours \cite{Pinel2006,Bardy2014,Zhao2015}. Despite previous attempts in the literature \cite{Qin2009}, the theory of similarity have not been tested in controlled experiments.

In this paper we demonstrate existence of a time-invariant {\it individual motor signature}, and show how it can be used to study socio-motor coordination. The notion of individual motor signature has its roots in the concept of frequency detuning (eigenfrequency difference) between two interacting humans and the phenomenon of the so-called maintenance tendency \cite{vonHolst1973,Rosenblum1988,Schmidt1994}. It has been further extended to intrinsic dynamics that is observed in form of individual preferred coordination modes (behavioural repertoire) exhibited during intra-personal coordination task \cite{Zanone1992,Kostrubiec2012}. 
We focus here on identifying a more general measure able to capture some key kinematic features of the motion of each individual and discriminate among different people. Using our measure, we are able to introduce a metric to assess the {\it dynamic similarity} between the movement of different humans and show that it helps to predict the level of coordination in an interactive joint-action task. 

We demonstrate the theory by using a virtual avatar playing the ``mirror game'', an activity where two players are asked to imitate each other's movements, and which has recently been established as a paradigm for studying interpersonal movement dynamics (see \cite{Hart2014,Noy2011} and references therein). Our evidence shows that similarity between the preferred motion of each player enhances the synchronisation level measured during the interaction.

More generally, our results introduce dynamic similarity as an important, complementary to qualitative measures of affiliation, factor affecting joint actions and enhancing coordination between socially interacting people.

\section{Methods}
\label{sec:methods}
The results presented in this paper are based on the analysis of data collected in three different experimental scenarios (each with different group of participants), performed in the course of the research project AlterEgo funded by the European Union \cite{alterego}. In Scenario~1 we only collected solo movements of the participants; in Scenario~2 we collected data from humans playing the mirror game in a solo condition and in dyads where they have to track each other's movements, in Scenario~3 we collected data from human participants playing solo and interacting with a virtual player (VP). Data collected in Scenario 1 were used to establish existence of the individual motor signature. Data collected in Scenarios 2 and 3 is used to demonstrate that coordination during an interactive task depends on the dynamic similarity between participants. In particular, we use Solo data collected in Scenarios 2 and 3 to measure dynamic similarity between interacting participants.

\subsection{Experimental Setups and Data Collection}
In Scenario 1, participants were asked to perform three solo sessions, each one separated by at least one week. Each participant was asked to sit comfortably on a chair and create interesting motion by moving her/his preferred hand above a leap motion$\circledR$ sensor  \cite{leapmotion} connected to a laptop. The movement of a participant was visualised on the screen of the laptop as a dot. Participants were given the following instruction: ``Play the game on your own, create interesting motions and enjoy playing''. Due to the nature of the experimental set-up, the position was recorded in arbitrary units. At each session, a participant was required to perform three solo rounds, each one lasting 60 seconds. In total, we recorded nine position time series for each of the 15 participants.

In Scenario 2 participants sat comfortably opposite each other. Two horizontal strings (length 1800mm) were mounted at eye level, centrally between the participants; on each string a ball with a small handle was mounted. Participants were instructed to move these balls left and right along the strings during the experiment. The movements of each participant were  captured using reflecting marker placed on the ball with infrared MX13 cameras (Vicon-Nexus, Oxford Metrics Ltd.) at a sampling rate of 100 Hz. Data was collected from 8 dyads (16 participants in total). All participants were right handed. Participants were given the following instructions:
\begin{itemize}
\item Solo Condition. Participants were given the same instruction as in Scenario 1. Participants had no view of their partner.
\item Leader-Follower Condition. ``This is a collaborative round whose purpose is to enjoy creating synchronised motion. Participant 1, lead the movement. Participant 2 try to follow your partner's movement.'' Two versions of this condition were played to allow both participants to lead and to follow.
\item Joint Improvisation Condition. ``In this collaborative round there is no leader and no follower. Let these 2 roles emerge naturally, imitate each other and create synchronised and interesting motions. Enjoy playing together.''
\end{itemize}

In Scenario 3, human players were asked to play with the VP described in \cite{Zhai2014cdc,Zhai2014smc}. Participants were standing in front of an LCD display showing the virtual player. A horizontal string (length 1800mm) was mounted in front of the participant. As in Scenario 2, a ball with a small handle was mounted on the string. Participants were instructed to move the ball left and right along the string. On the screen facing the human player, a ball, which is controlled by the VP, is also shown to move along a string. The movement of each participant was recorded with a single wide-angle camera. The sampling rate was not uniform and averaged around 40Hz. 

The virtual player was driven by an interactive cognitive architecture (ICA) which used a pre-recorded reference motion trajectory ($\mathrm{Ref}$) and an adaptive feedback control algorithm to generate the VP's movement, while being influenced by the follower's performance (see \cite{Zhai2014cdc,Zhai2014smc} for further details). It is important to note that the ICA does not simply replay pre-recorded time-series as in \cite{Noy2015}, but uses them as preference signal in order to generate the output trajectory for the VP. This allows for a real-time movement behaviour matching between human and virtual players, which is a fundamental part of the interaction in the mirror game. For instance, if the follower stops tracking the movement of the leader, it is appropriate for the leader, as done by the ICA driving the virtual player, to adjust its movement and guide the follower in order to encourage the interaction. In Scenario 3 the interactive cognitive architecture driving the VP was fed with pre-recorded position time series based on solo trials of the human participant playing with it. More specifically, to control similarity between the solo motion and reference trajectory the pre-recorded solo trajectory of the player was superimposed with a 2.5Hz sinusoidal signal with time varying amplitude defined as $1/3$ of the corresponding normalised velocity of the solo trajectory. Further analysis of the VP's performance can be found in {\it SI appendix}. Data was collected from 51 individuals playing the mirror game with the virtual player. The data set of each participant contained participant's and VP's positions for each of the following rounds: 4 solo (one minute) rounds (without VP) and 12 rounds (30 seconds) where the human participant played as a follower.

\subsection{Data processing}
The collected data was pre-processed in Matlab$\circledR$. When necessary we used interpolation with shape-preserving piecewise cubic interpolation and filtering with a zero-phase forward and reverse digital 2nd order lowpass (10Hz cut-off) Butterworth filter. The position time series were then used to numerically estimate their corresponding velocity time-series. To differentiate position time-series we used a fourth-order finite difference scheme. We cut out the first and last 2 seconds of the signal. Furthermore, we limited velocities to 3.5 [a.u./s] in the experimental Scenario 1 and to 2.7 [m/s] in the experimental Scenarios 2 and 3 (higher velocities were considered as results of noise in the collected data). To estimate the probability density function (PDF) of the player's velocity we use normalised histogram of the velocity time series with 101 equally distant bins between -2.7 and 2.7 [m/s] (or -3.5 and 3.5 [a.u./s] in Scenario 1). Further details about data processing can be found in  {\it SI appendix}.

In order to quantify similarity between PDFs of player's velocity we use Earth's movers distance (EMD) that is an established tool in pattern recognition applications \cite{Levina2001}. Intuitively, the EMD measures how much work is required to transform a 'pile of earth' into another; each 'pile of earth' representing a histogram. In the case of univariate probability distributions the EMD is given by the area of the difference between their cumulative distribution functions. More details can be found in {\it Appendix A}.

Distances between PDFs are then analysed by means of Multidimensional scaling (MDS). MDS is a well established tool in data visualisation and data mining \cite{Borg2005}. It allows to reduce dimensionality of the data and visualise relations between the objects under investigation while preserving as much information as possible. Since the EMD is a metric in the space of the PDFs of velocity time series we use classical MDS in the form implemented in Matlab$\circledR$. Description of the MDS algorithm and discussion of relation between MDS and principal component analysis can be found in {\it SI appendix}.

In order to quantify temporal correspondence (level of coordination) between players we introduce the {\it relative position error} (RPE). The RPE is a measure of temporal correspondence between complex, non-periodic, coordinated movements which is based on the natural notion of a follower lagging behind the leader when tracking her/his motion. In particular, RPE is a measure of position mismatch between a leader and a follower capturing how well the follower tracks the leader's movement. See {\it Appendix C} for further details. Discussion of the advantages of using the RPE over the relative phase based on the Hilbert transform can be found in Sections 9 and 10 of the {\it SI appendix}.

\section{Results}
\label{sec:results}

We show the existence of an {\it individual motor signature} (IMS) for each player, defined as a time-invariant tractable characteristic of her/his movement. We develop a framework allowing us to demonstrate that characteristics of the solo movement in the mirror game are time-persistent and differ significantly between participants. In the proposed framework we employ  {\it velocity profiles} (PDFs of the velocity time series) to reveal that the rapport (similarity) between individual motor signatures enhances synchronisation of movement between participants in joint action. Finally, we demonstrate that a virtual player driven by a novel interactive cognitive architecture \cite{Zhai2014cdc,Zhai2014smc} can be used to study interpersonal interaction in a 'mirror game' between human and virtual players.

\begin{figure}[t!]
 \centering
 \includegraphics[width=\textwidth]{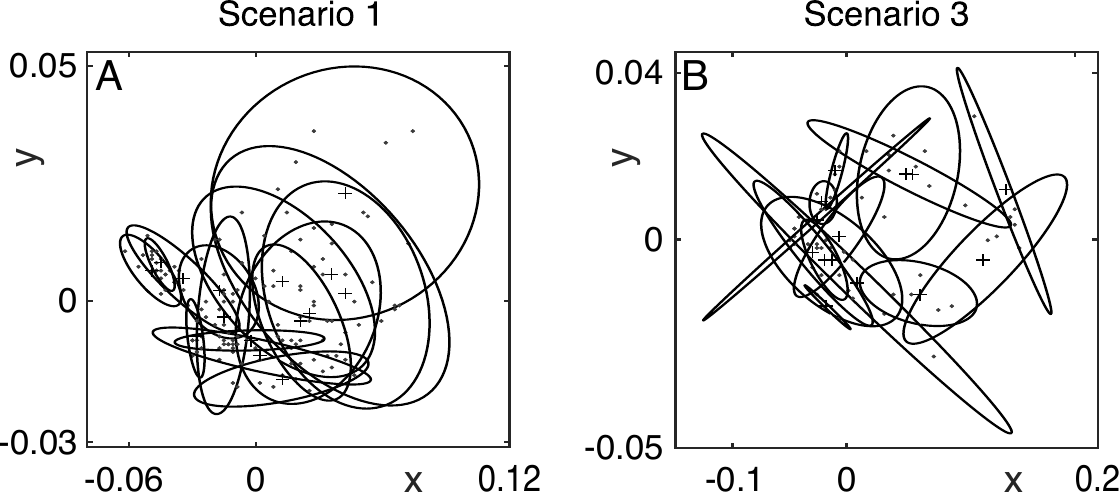}
 \caption{Individual motor signature in the {\it similarity space} computed with MDS from distances between velocity profiles. (A) For 15 different participants from solo mirror game recordings in Scenario 1, on three different days with at least one week break between recording rounds. (B) For 56 solo trials of 14 participants from solo mirror game recordings in Scenario 3 (for the sake of clarity data for only 14 out of 51 participants is shown). Each ellipse corresponds to a different participant. Small dots correspond to individual solo recordings. Each cross at the centre of an ellipse corresponds to the average of the small dots' positions. Each ellipse indicates 0.7 mass of bivariate normal distribution fitted to the small dots (see {\it SI appendix} for further details).}
 \label{fig:IMS_vel_prof}
\end{figure}

\subsection{Existence of an Individual Motor Signature}
\label{sec:IMS}

In this section we study solo mirror game recordings collected in the experimental Scenario 1 in order to investigate the existence of an {\it individual motor signature}. In particular we demonstrate that: 1) the movement characteristics of each individual persist in time, and 2) that they differ significantly between individuals.

To this end, we analyse {\it velocity profiles} 
which characterise motion in the mirror game on the time-scale of a complete experimental trial. We use the Earth Mover's Distance (EMD) to assess distances between velocity profiles of different individuals. We then represent them as points in the {\it similarity space}, that is an abstract geometric space constructed by means of multidimensional scaling (MDS) \cite{Borg2005} that provides a visual representation of the pattern of proximities (i.e., similarities or distances) among a set of objects. As a result we obtain clusters of points corresponding to solo trials of individual participants. In order to measure separation between clusters of points in the similarity space we measure overlap $\omega$ between ellipses that encircle them, with $\omega=0$ meaning that the ellipses do not overlap at all and $\omega=1$ meaning complete overlap (see {\it SI appendix} for further details about ellipses and overlap). 

Figure~\ref{fig:IMS_vel_prof} depicts velocity profiles from solo trials presented as elements of the {\it similarity space}. Figure~\ref{fig:IMS_vel_prof}A shows data for 15 different participants from experimental Scenario 1 and Fig.~\ref{fig:IMS_vel_prof}B  shows representative data of 14 out of 51 participant from experimental Scenario 3. We note that, in the experimental Scenario 3, players had larger range of movement and all the solo trials of individual players were recorded on a single day. Each dot in Fig.~\ref{fig:IMS_vel_prof} corresponds to a velocity profile from a single trial. The dots corresponding to different individuals are encircled by ellipses.
Importantly, Fig.~\ref{fig:IMS_vel_prof}A demonstrates clustering of the dots for different participants collected on three different days. Such clustering indicates time-invariance of the individual motor signature. The variability between radii of the ellipses associated with different individuals signify that the individual motor signature of some individuals is more variable than the others. 

Furthermore both data sets presented in Fig.~\ref{fig:IMS_vel_prof} demonstrate a good separation of the ellipses corresponding to individual participants, with a median overlap $\omega$ between 15 ellipses in A equal to 0.02 and between ellipses of all 51 participants from experimental Scenario 3 equal to 0.05. Interestingly, in the data from experimental Scenario 1 there are 45 out of 105 pairs of ellipses that do not overlap at all, meaning that in almost half of the cases two participants can be explicitly distinguished just by observing their solo motion. The same holds true for 418 out of 1275 pairs of ellipses from experimental Scenario 3. Comparison of separation between individuals achieved by means of the velocity profiles and using individual characteristic of motion suggested in \cite{Hart2014,Noy2011,Noy2015} can be found in the {\it SI appendix}.

The physical interpretation of the two principal dimensions of the similarity space constitutes a further insight gained from our analysis. In particular, our analysis reveals that the coordinate $x$ of the movement representation in the similarity space, which corresponds to the first principal dimension given by the MDS, is correlated with the absolute average velocity of the motion. In addition the $y$ coordinate of the representation of each time series in the similarity space, which corresponds to the second principal dimension from the MDS, is correlated with the kurtosis of a velocity segment \cite{Hart2014,Noy2011}, a part of the velocity time-series between two consecutive times of zero velocity. That is it informs us on the ratio of high and low velocities in the motion. For further details about interpretation of the dimensions in the similarity space see {\it Appendix B}.

In summary, individual motor signatures identify each different participant and can be used effectively to measure dynamic similarity between them. More importantly, they provide a comprehensive and holistic description of the kinematic characteristics and variability of human movement. 

\subsection{Behavioural plasticity during social interaction}
\label{sec:Plasticity}

\begin{figure}[t!]
 \centering
 \label{fig:cond}
 \includegraphics[width=0.75\textwidth]{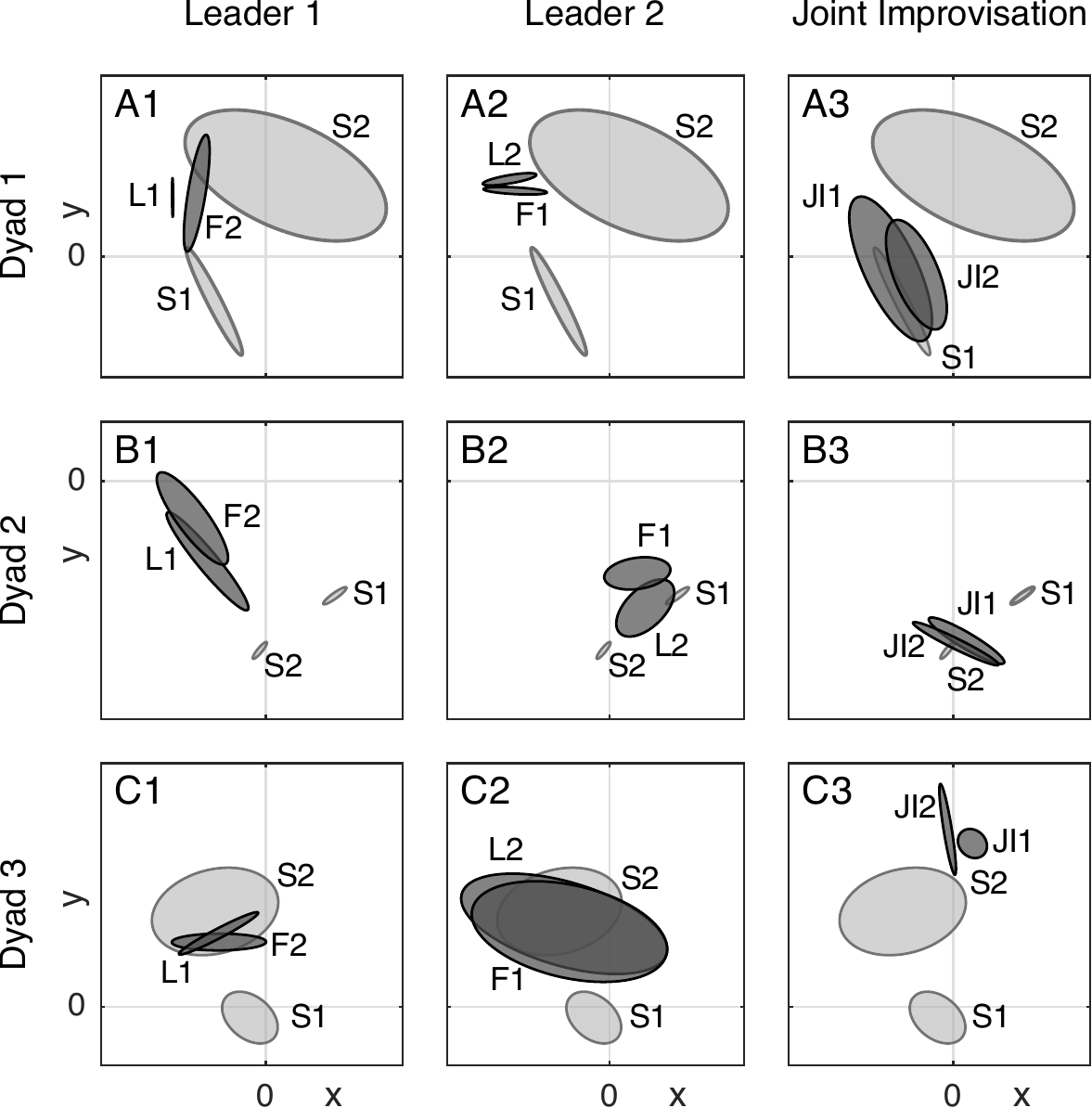}
 \caption{Interaction between two players in different experimental conditions visualised in the {similarity space}. Ellipses encircle points corresponding to velocity profiles in solo (S1 and S2; light grey), leader (L1 and L2; dark grey), follower (F1 and F2; dark grey) and joint improvisation (JI1 and JI2; dark grey) rounds. Each row depicts data for a different dyad. In column 1 player 1 was a leader, in column 2 player 2 was a leader and in column 3 participants played in joint improvisation condition. $x$ axis has the same range in all panels, $y$ axis is rescaled for clarity of presentation.}
\end{figure}

Using the concepts of {\it individual motor signature} and {\it similarity space} we are able to demonstrate behavioural plasticity during social interaction. Specifically by behavioural plasticity we mean that in order to cooperate people are willing, to a different degree, to disregard their individual preferences. Indeed, by comparing positions in the {similarity space} of the velocity profiles from solo and cooperative trials (leader follower and joint-improvisation), we find that some people are more inclined to adjust their kinematic characteristics when interacting with others in the mirror game. Figure~2 shows three representative examples of the behavioural plasticity detected during the experimental Scenario 2. 
In Fig.~2A1 and A2 we depict consistent behaviour of the leader and follower independently of which player is the designated leader. In Fig.~2B1 and B2 we notice that dynamics of movement differs significantly depending on who is leading. In Fig.~2C1 and C2 we illustrate a player (S2) that dominates the interaction in terms of movement characteristics. Furthermore, Fig.~2A3--C3 show that motion dynamics in the joint improvisation condition is clearly different compared to the leader-follower condition in agreement with recently published results \cite{Hart2014}. Visualisation of the interactions for all the 8 dyads from the experimental scenario 2 can be found in {\it SI appendix} Supplementary Figure 4. Taken together the above observations suggest that the behavioural plasticity might be regulable rather than fixed and could be modulated in order to enhance social competence. Fig.~2 showcases a new technique that could be useful in studies of mutual rapport, affiliation and leadership emergence. However, rigorous analysis of the complex relation between individual motor signatures and movements during cooperative conditions is beyond the scope of this paper.

\subsection{Dynamic similarity enhances coordination in joint action}
\label{sec:tc_ds}

\begin{figure}[t!]
 \centering
 \includegraphics[width=0.75\textwidth]{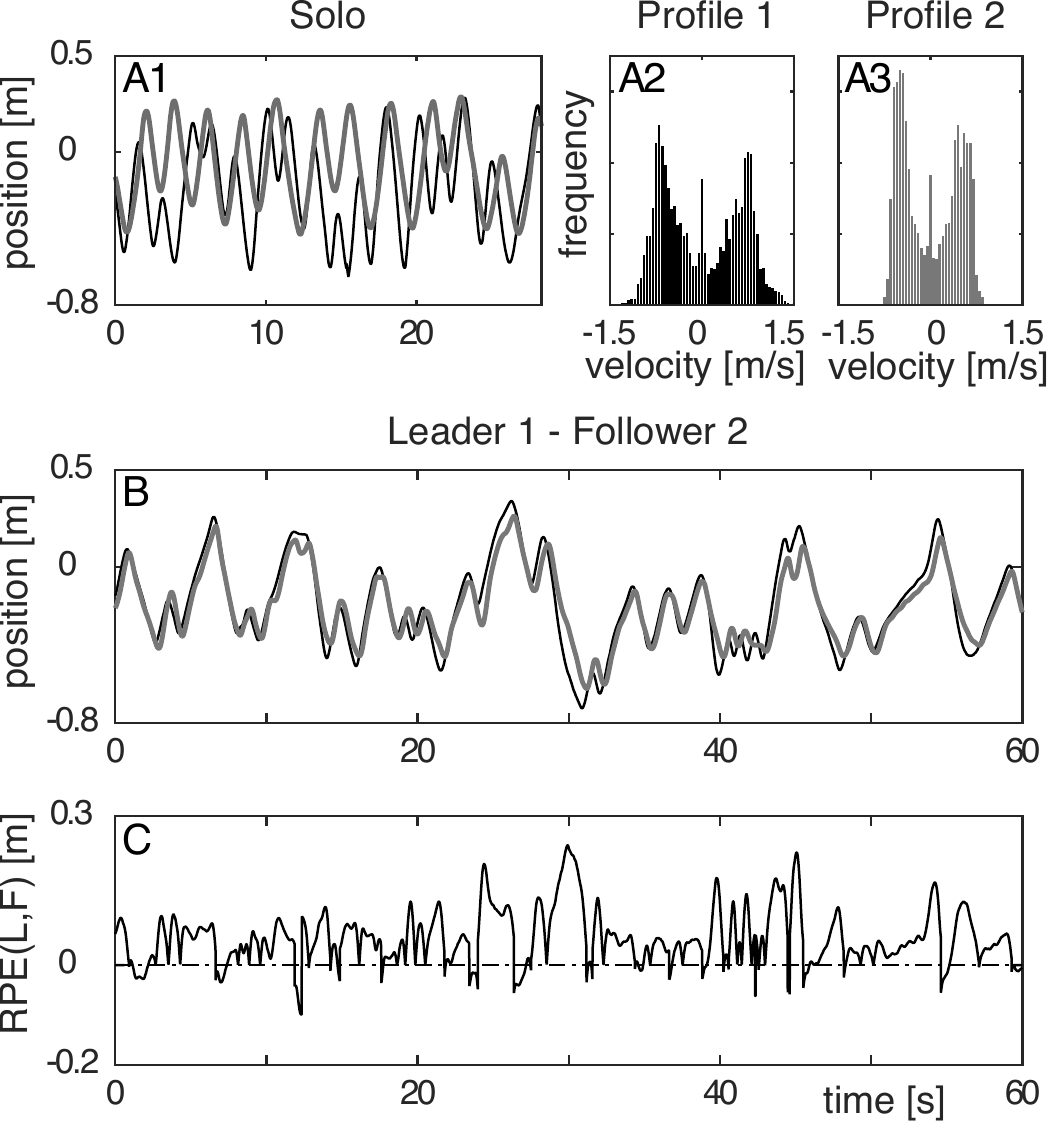}
  \caption{First we compute dynamic similarity between two players. In panel A1 we show the solo movements of two participants who later played together in the leader-follower condition. Panels A2 and A3 depict velocity profiles that represent individual motor signatures of the two players (S$_{A2}$ and S$_{A3}$) corresponding to the positions time series presented in panel A1. The EMD(S$_{A2}$,S$_{A3}$) = 0.0303 between the histograms in panels A2 and A3 quantifies dynamic similarity between the two players. Then we measure temporal correspondence between their movements when they play together in the leader-follower condition. Panel B illustrates position traces of the participants from panel A when they play together as a leader (black) and follower (grey). Panel C shows the RPE between leader and follower trajectories presented in panel B. The mean value and the standard deviation of the RPE are respectively $\mu \mbox{RPE}$(L$_{A2}$,F$_{A3}$) = 0.05 and $\sigma \mbox{RPE}$(L$_{A2}$,F$_{A3}$) = 0.05.}
\label{fig:method}
\end{figure}

Having defined suitable measures, we next analyse the effects of the dynamic similarity on the temporal correspondence in two different experimental scenarios: the former where two humans play the mirror game (Scenario 2) and the latter where a human is asked to play the game with a virtual player (Scenario 3). In particular, we measure the temporal correspondence between players in the leader-follower condition of the mirror game and study systematically if and how it is related to the difference between their individual motor signatures. 
\begin{figure}[t!]
 \centering
 \includegraphics[width=\textwidth]{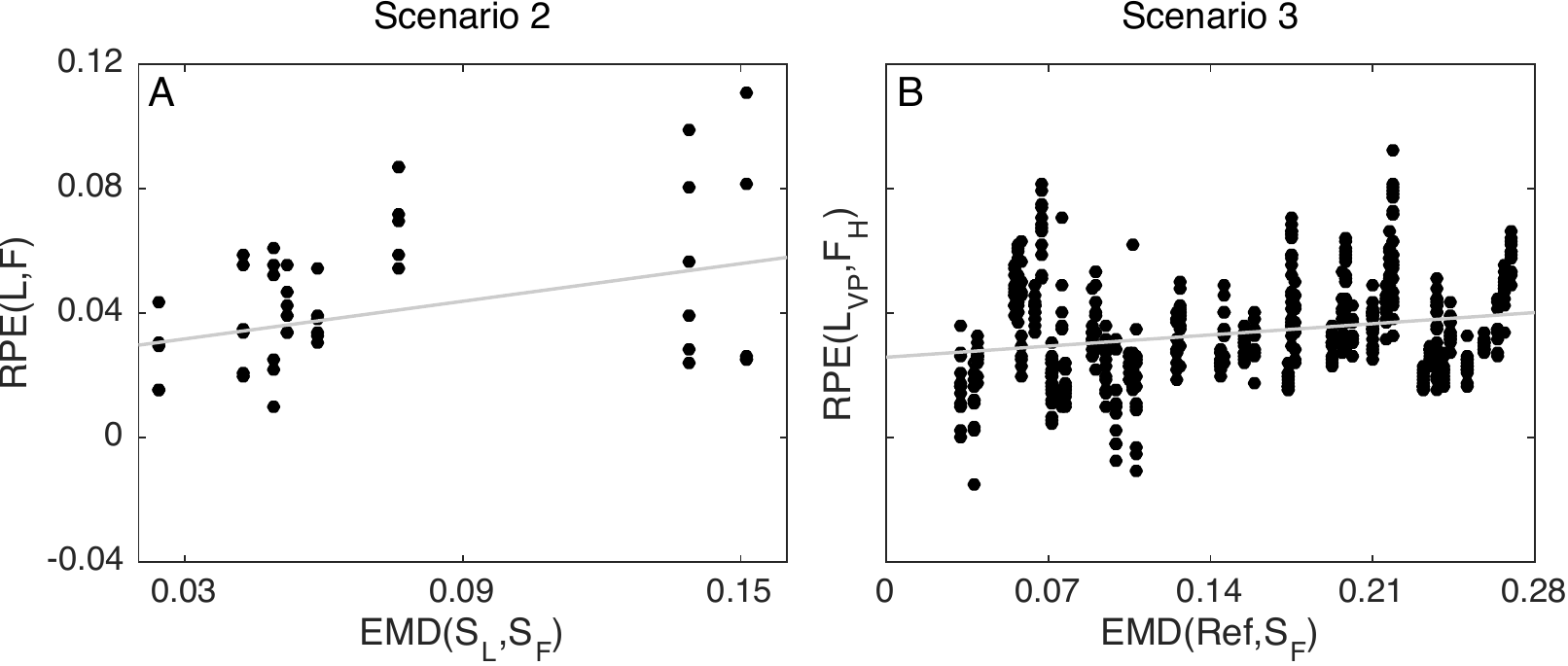}
  \caption{Panel A shows correlation between EMD(S$_\mathrm{L}$, S$_\mathrm{F}$) and RPE(L, F) computed for all individual leader-follower trials in 8 dyads from Scenario 2. Panels B depicts the dependence of RPE(L$_\mathrm{VP}$, F$_\mathrm{H}$) between the VP leading the human participant on EMD(Ref, S) between the reference trajectory and the participant's solo movement (Scenario 3). Each black dot corresponds to a single leader-follower trial. Grey lines are presented only for illustrative purposes. Spearman's $\rho$ coefficients are equal to: $\rho_A=0.3907$ $(p_{\rho A}=0.0066)$, $\rho_B=0.2224$ $(p_{\rho B}<1.0e-5)$; Pearson's $R^2$ coefficients are equal to: $R^2_A=0.3701$, $(p_{R^2 A}=0.0105)$, $R^2_B=0.2343$ $(p_{R^2 B}<1.0e-5)$.}
\label{fig:corr}
\end{figure}
In order to demonstrate that dynamic similarity facilitates coordination between players in the mirror game we seek to find a correlation between dynamic similarity, as quantified by means of the distance between velocity profiles in the similarity space, and temporal correspondence measured by the relative position error (RPE).  For further details about the definition and interpretation of the RPE see {\it Appendix C}.
Fig.~\ref{fig:method}A--C illustrate the steps we take in our analysis detailed in the figure legend. 

The correlation between temporal correspondence and dynamic similarity observed in the data from human-human interaction collected in the experimental Scenario 2 is shown in Fig.~\ref{fig:corr}A. For each dyad we calculate nine values for the distance between the players signatures S$_\mathrm{L}$ and S$_\mathrm{F}$ (all the combinations between three solo trials for each player) and six values for the mean relative position error between the leader (L) and the follower (F) (three trials with player 1 as a leader and three trials with player 2 as a leader); Spearman's rank correlation \cite{Hogg2005, Corder2014} between EMD distance and relative position error was computed to be $\rho$=0.3907 ($p_{\rho}$=0.0066). We removed a single outlier with RPE$>$0.2 (the correlation including the outlier was stronger). We use Spearman's rank correlation because our data is not normally distributed. 

We further control the identified correlation for two confounding factors. First, we expect that a faster leader is more difficult to follow than a slower one, hence we control the identified correlation for the mean absolute velocity of the leader $\mu|V_L|$. Second, we expect that it is easier for a follower who prefers fast motions to track movement of a leader who prefers to move slowly, than it is for a follower who prefers slow motions to track a leader who prefers to move fast. Therefore we control the identified correlation for the difference between mean absolute velocities of the solo movements of the leader and the follower  $\mu|V_{SL}|-\mu|V_{SF}|$, which is proportional to the difference between their $x$ coordinates (see {\it Appendix B}). Partial and adjusted correlation coefficients are computed in Matlab$\circledR$ using functions {\tt partialcorr} and {\tt partialcorri}. As expected, we found the mean RPE between leader (L) and follower (F) to be strongly correlated with both $\mu|V_L|$ and $\mu|V_{SL}|-\mu|V_{SF}|$. Nonetheless, the relation between mean RPE(L, F) and EMD(S$_\mathrm{L}$, S$_\mathrm{F}$) controlled for the confounding factors remained statistically significant; partial Spearman's rank correlation controlled for $\mu|V_L|$ was computed to be $\rho$=0.3466 ($p_{\rho}$=0.0183);  partial Spearman's rank correlation controlled for $\mu|V_{SL}|-\mu|V_{SF}|$ was computed to be $\rho$=0.5025 ($p_{\rho}$=0.0003); Spearman's rank correlation adjusted for both, $\mu|V_L|$ and $\mu|V_{SL}|-\mu|V_{SF}|$, was computed to be $\rho$=0.4697 ($p_{\rho}$=0.0011).

We confirm our observations by also analysing data from the experimental Scenario 3, where a virtual player leads a human follower. In particular, we compute correlations between the similarity of the two players' motions (evaluated in terms of the average distance between the velocity profile of the VP reference trajectory ($\mathrm{Ref}$) and the four velocity profiles of the solo motion of the human player) and their temporal correspondence.
Fig.~\ref{fig:corr}B demonstrates that temporal correspondence depends on the dynamic similarity between the reference trajectory of the virtual player and participant's solo movement. In particular, we find that mean of the RPE between leader (L$_\mathrm{VP}$) and follower (F$_\mathrm{H}$) increases with the distance between their signatures. This finding affirms that dynamic similarity between reference trajectory and player's solo movements facilitates coordination between the VP leader and human (H) follower; the Spearman's rank correlation was computed to be $\rho$=0.2224 ($p_{\rho}<$ 1e-05). 

As in the case of the data from the experimental Scenario 2 we controlled the correlation for the two confounding factors and found that the partial correlation controlled for $\mu|V_{L_{VP}}|$ is significant with partial Spearman's rank coefficient equal to $\rho$=0.1863 ($p_{\rho}<$ 1.0e-05). On the other hand, the correlation disappears when we control it for $\mu|V_{Ref}|-\mu|V_{SF}|$. However, since the reference trajectories of the VP were simply made faster by adding 2.5Hz sine signal, i.e. $\mu|V_{Ref}|\geq\mu|V_{SH}|$, this is exactly what should be expected. In other words, our analysis of the data from the experimental Scenario 3 demonstrates the effect of the dynamic similarity in the special case when it can be simplified to a difference between the preferred solo velocities of the participants. It is important to note that this observation is only possible due to our experimental set-up allowing for an interaction between a human and a virtual player using as a reference trajectory one of the human player's solo trajectories post-processed as described in the Methods Section above.

In summary, we show that a small distance between individual velocity profiles of the leader and the follower, indicating that they have similar movement dynamics, results in higher levels of coordination than those observed in dyads in which the distance between participants' individual motor signatures is larger. In so doing, we demonstrate that dynamic similarity affects the level of coordination in joint human movement interactions. Our results are a step forward towards confirming a prediction of the theory of similarity, namely that dynamic similarity enhances inter-human interaction.

\section{Discussion and Conclusions}
In this paper we introduced the notion of dynamic similarity in the mirror game \cite{Noy2011} and demonstrated the existence of an individual motor signature in human players. We then  
showed that the synchronisation level between the players is affected by their dynamic similarity.  
In particular, we proposed the use of velocity profiles, defined by the probability density functions of velocity time series recorded in the mirror game, as motor signatures.  We used the earth's mover distance and multidimensional scaling to show that velocity profiles of the solo movement have characteristics of individual motor signatures, i.e. they are stable over time and differ significantly between individual players. In this way we revealed time-persistent, individual motor properties that could be detected in complicated, non-periodic motion observed in the mirror game, as suggested in \cite{Hart2014}.  Notably we extended the notion of motor signature beyond the frequency content of a periodic motion \cite{vonHolst1973,Rosenblum1988,Schmidt1994,Issartel2007}. 
Since the individual motor signature can be readily recorded by means of a cheap off-the-shelf experimental set-up, we believe that it could become an integral part of studies investigating interpersonal interaction.

We introduced the evaluation of the distance between velocity profiles as a method of quantifying dynamic similarity between players' motion in the mirror game and used the relative position error to measure their temporal correspondence.
Our key finding supports a central predictions of the theory of similarity, specifically that dynamic similarity of participants' solo motions enhances their coordination level \cite{Schmidt1994,Varlet2014}.

Our work complements research on individuality and interactions in animal groups in two ways \cite{Herbert-Read2013}. First, our study involves direct and intentional coordination that is typical for human-human interactions. Such interactions, in general, allow for investigation of an intentionally designated leader's behaviour and are fundamentally different compared to spontaneous or unintentional coordination amongst individuals and/or groups of animals. Second, the overlap between our results and the findings reported in \cite{Herbert-Read2013} represents a promising avenue for future work on the extension of animal models to inter-personal interactions.

Finally, the methods we have introduced and the data we have collected establish the use of a virtual player \cite{Kelso2009}, driven by an interactive cognitive architecture, as an effective tool for studying joint actions in the mirror game. 
Importantly, the advantages of using an interactive cognitive architecture based on feedback control to drive the avatar is that bi-directional coupling is maintained during the mirror game and that it allows control of the interaction between human and avatar in two ways, by choosing reference trajectories and by changing the bi-directional coupling parameters of the interactive cognitive architecture. Such level of control in a socio-motor coordination task could be used in applications that aim to reinforce social bonding in joint-action tasks \cite{Oullier2008,Dumas2014}.

In summary:
\begin{itemize}
\item We introduce quantitative measures and analyse dynamic properties of complex aperiodic movements that characterise human socio-motor interactions.
\item We demonstrate the existence of an individual motor signature for each player, defined as a time-invariant tractable characteristic of her/his movement and reveal that the rapport (similarity) between individual motor signatures enhances coordination of movement between different players.
\item We introduce a novel interactive cognitive architecture able to drive a virtual player to play the game and employ it as a model of interpersonal interaction in the mirror game between human and virtual players.
\end{itemize}

\section*{Competing interests}
The authors declare no competing financial interests.

\section*{Author Contributions}
P.S., C.Z., F.A., R.S., M.G., L.M., B.G.B., M.dB., K.T-A. conceived the study and designed the experiments; C.Z., F.A., R.S., M.G., and L.M. performed the experiments; B.G.B and L.M. administered the experiments; P.S. developed analytical tools, carried out data analysis and wrote the paper; M.dB. and K.T-A. supervised the analysis and wrote the paper. All authors discussed the results and implications and commented on the manuscript at all stages. All authors gave final approval for publication.

\section*{Acknowledgments}
The authors would like to thank Ed Rooke (University of Bristol), Catherine Bortolon (University Hospital of Montpellier) and Zhong Zhao (University of Montpellier I) for help in data collection, as well as preparing and running some of the experiments. We would also like to thank Richard Schmidt and Michael Richardson for helpful discussions. We are also grateful for the insightful comments of the anonymous reviewers.

\section*{Funding}
This work was funded by the European Project AlterEgo FP7 ICT 2.9 - Cognitive Sciences and Robotics, Grant Number 600610. The research of K.T-A was supported by grant EP/L000296/1 of the Engineering and Physical Sciences Research Council (EPSRC). 

\section*{Appendix A}
\subsection*{Earth's movers distance} 

\begin{figure*}[t!]
 \centering
 \includegraphics[width=\textwidth]{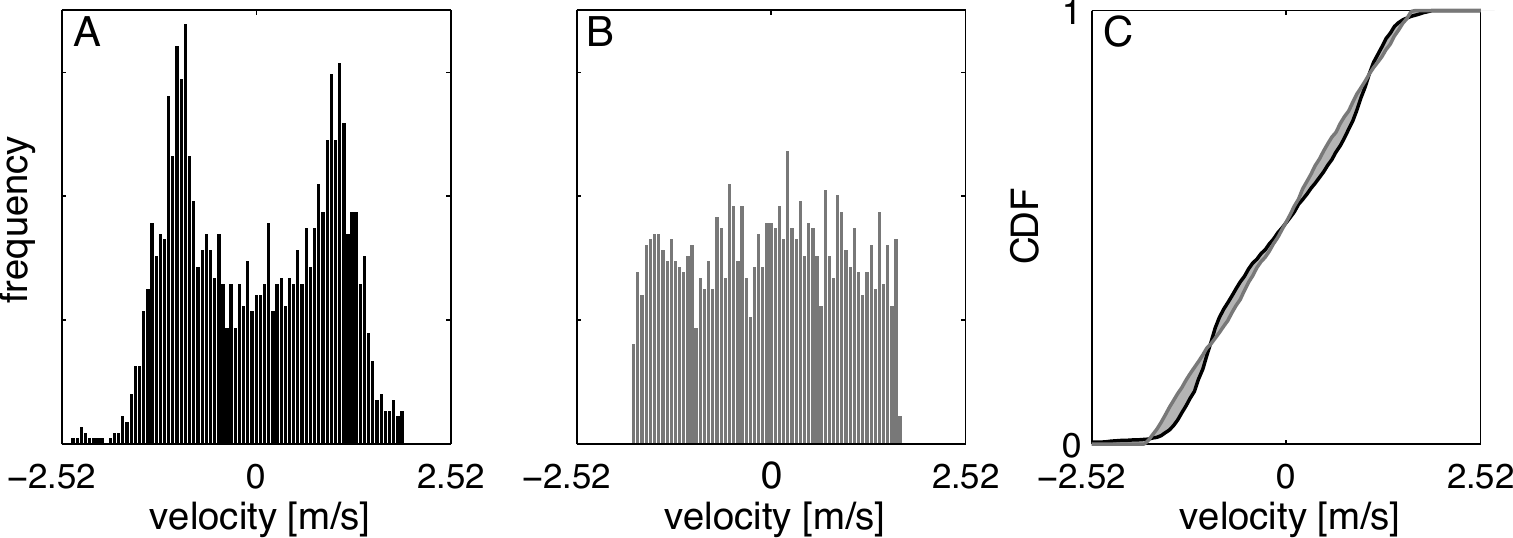}
 \caption{Panel A shows a velocity profile - histogram ($h_A$; black) of the velocity time series. Panel B shows histogram ($h_B$; grey) of a random variable generated with distribution of type-1 from the Pearson system. Both samples have $\mu=-0.01$, $\sigma=0.54$, $s=-0.04$ and $k=1.82$ where $\mu$ is the mean, $\sigma$ is the standard deviation, $s$ is the skewness and $k$ is the kurtosis. Panel C shows cumulative density functions of the distribution from panel A in black and from panel B in grey. The light grey shading indicates the area of difference between the CDFs, i.e. the Earth's movers distance between the two distributions, $EMD(h_A,h_B)=0.02$.}
 \label{fig:k_vs_k}
\end{figure*}

Mathematically the Earth's movers distance (EMD) is a special case of Wasserstein distance \cite{Levina2001}. It is a solution of the optimal transportation problem \cite{Kantorovich1958} and is an established tool in image analysis and pattern recognition applications \cite{Levina2001}. For univariate PDFs it can be expressed in a closed form as the area between their corresponding cumulative distribution functions \cite{Cohen1997}:
\begin{equation}
EMD(PDF_1(z),PDF_2(z))=\int_Z|CDF_1(z)-CDF_2(z)|dz
\end{equation}
Here, $PDF_1$ and $PDF_2$ are two probability density functions with support in set $Z$, $CDF_1$ and $CDF_2$ are their respective cumulative distribution functions. We note that EMD is a well-defined metric in the space of PDFs as it satisfies the following conditions:
\begin{align*}
\mbox{Positive definiteness: }&d(x_1,x_1)=0,\,x_1 \neq x_2 \Rightarrow d(x_1,x_2)>0, \\
\mbox{Symmetry: }&d(x_1,x_2)=d(x_2,x_1),\\
\mbox{Triangle inequality: }&d(x_1,x_2)\leq d(x_1,x_3)+d(x_2,x_3).
\end{align*}
These conditions express intuitive notions about the concept of distance. For example, that the distance between distinct points is positive and the distance from $x$ to $y$ is the same as the distance from $y$ to $x$. The triangle inequality means that the distance from $x$ to $z$ via $y$ is at least as great as from $x$ to $z$ directly. Furthermore, EMD is non-parametric and quantifies partial matches. Hence, it allows to compare PDFs rather than their selected moments. This represents a significant advantage of our analysis in comparison to using selected moments, which are not sufficient to uniquely parameterise a PDF. We note that a bounded PDF is uniquely determined by its moments of all orders (0 to infinity) \cite{Billingsley2012}.

Figure~\ref{fig:k_vs_k} shows an example of two histograms with the same estimates of the first four moments and demonstrates that by using the EMD we can distinguish them. The histogram in panel A is a velocity profile ($h_A$) computed for a time series in our data, while panel B shows a histogram ($h_B$) of a random variable generated using distribution of type-1 from the Pearson system \cite{Johnson1994}. Figure~\ref{fig:k_vs_k}C illustrates how the EMD between two experimental PDFs is computed. The black line in Fig.~\ref{fig:k_vs_k}C is the experimental CDF of the histogram from panel A, while the grey line is the experimental CDF of the histogram from panel B. The $EMD(h_A,h_B)$ between the two velocity profiles is the area between the black and the grey line, which is indicated with the light grey shading. We note that the maximum of the difference between two CDFs is a basis of the non-parametric goodness-of-fit Kolmogorov-Smirnov test \cite{Chakravarty1967}. 

In practice, to compute empirical CDFs we take the cumulative sum of histogram bins normalised by the number of data points. We use histograms with 101 equidistant bins over fixed range of velocity values (outliers are assigned to the most extreme bins). Furthermore, we normalise the EMDs with the maximal EMD for a given support $Z$; since $|CDF_1(z)-CDF_2(z)| \leq 1$, the maximal EMD is given by the length of the support $EMD_{max}=|Z|$. 

\section*{Appendix B}
\subsection*{Similarity space} 

\begin{figure*}[t!]
 \centering
 \includegraphics[width=11cm]{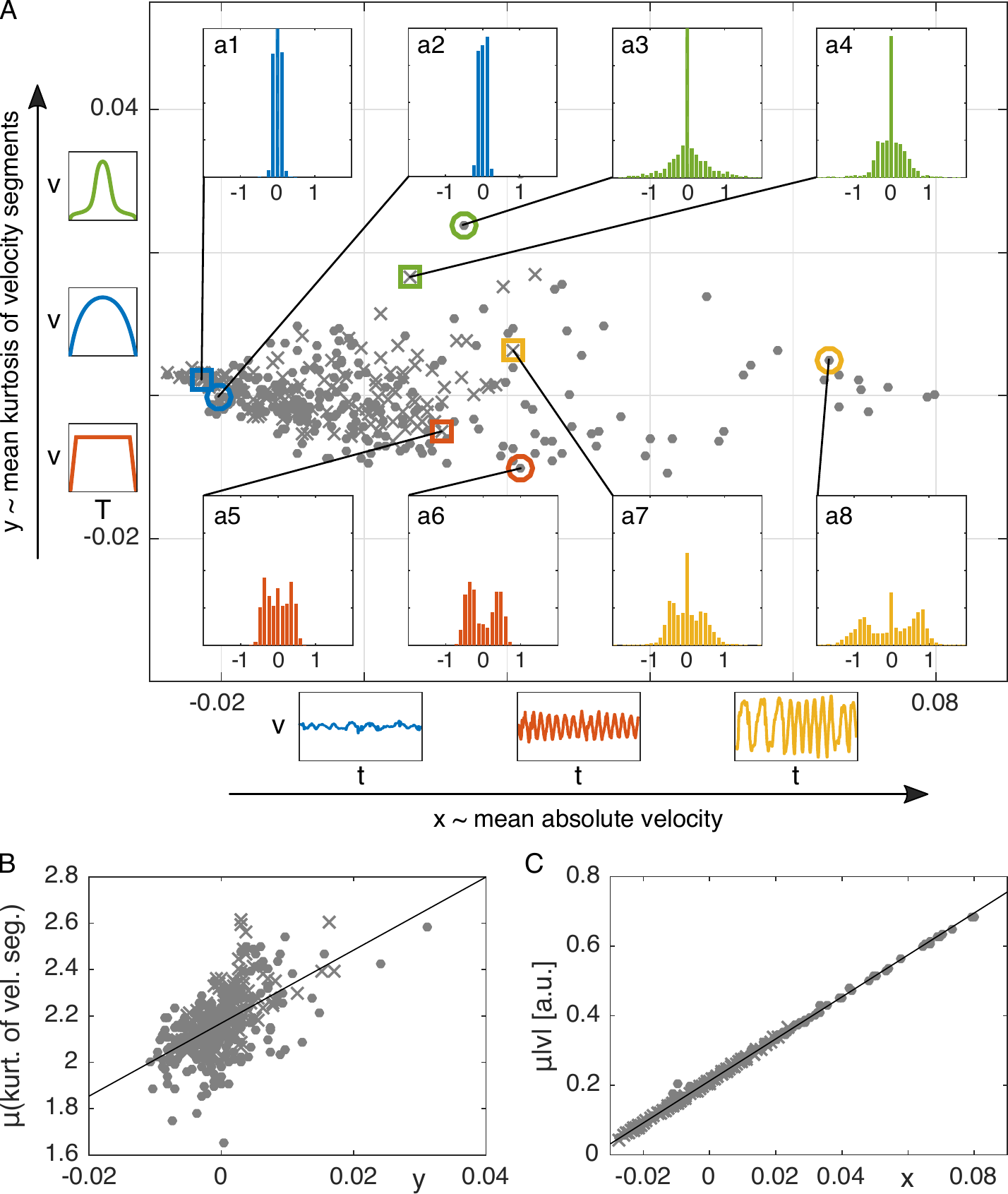}
 \caption{(A) {\it Similarity space} showing points corresponding to velocity profiles of normalised solo data from the Scenarios 1 (x) and 3 (\textbullet). Insets (a1)--(a8) show examples of the velocity profiles.  $v$ is velocity, $t$ is time and $T$ indicates normalised duration of the velocity segment. (B) Relation between $y$ coordinate of the similarity space and average kurtosis of the velocity segments: Corr($y$, $\mu$(kurtosis of vel. segments)): $R^2$=0.5657 ($p_{R^2}$=0), $\rho$=0.5685 ($p_{\rho}$=0).
 (C) Relation between $x$ coordinate of the similarity space and average solo velocity of a participant: Corr($x$,~$\mu|v|$): $R^2$=0.9993 ($p_{R^2}$=0), $\rho$=0.9979 ($p_{\rho}$=0).
 }
 \label{fig:sim_space}
\end{figure*}

In Fig.~\ref{fig:sim_space} we illustrate properties of the {similarity space}. Figure~\ref{fig:sim_space}A shows points corresponding to velocity profiles of solo data from Scenario 1 (x) and solo data from Scenario 3 (\textbullet). In order to show the points from two different experimental set-ups in one plot we normalised position time-series before computing velocities. Additionally, since the range of motion in the Scenario 1 is equal to 60cm and in the Scenario 3 it is equal to 180cm, we multiplied normalised position data from the Scenario 1 by $1/3$.  Insets (a1), (a4), (a5) and (a7) show examples of velocity profiles from the the experimental Scenario 1 and (a2), (a3), (a6) and (a8) show examples of velocity profiles from the the experimental Scenario 3.

Further analysis of the data revealed that the $x$ coordinate of the similarity space is correlated with mean absolute velocity $\mu|V|$, see Fig.~\ref{fig:sim_space}C, and that $y$ coordinate of the similarity space is correlated with mean kurtosis of normalised velocity segments, see Fig.~\ref{fig:sim_space}B, parts of the velocity time-series between two consecutive points of zero velocity. Mean kurtosis of the velocity segments describes how participant is changing direction of motion. Low kurtosis indicates sudden changes of directions and relatively constant velocity between them, while high kurtosis informs us that the change of direction was slow. The two quantities, taken together, explain accurately general characteristics of the velocity profiles, e.g. bi-modality of the velocity profile in the inset (a6) means that the participant was moving with a high constant velocity and was quickly changing direction of motion. On the other hand, velocity profile in the inset (a3) tells us that the participant was changing the direction slowly, high peak close to zero velocity, and was reaching maximum velocity only for brief moments.

More generally, Fig.~\ref{fig:sim_space} shows that the dimensions of the {similarity space}, computed using multidimensional scaling of earth's mover distances between velocity profiles, emerge from properties and characteristics of the human motion in the mirror game. It is important to note that a single property of the velocity profiles, say $\mu|V|$, is not sufficient to separate different participants. Furthermore, we observe better separation between individuals using the coordinates of the {similarity space} than using a plane defined by their correlates.

\section*{Appendix C}

\subsection*{Relative position error}
\label{sec:RPE}

\begin{figure}[t!]
 \centering
 \includegraphics[width=\textwidth]{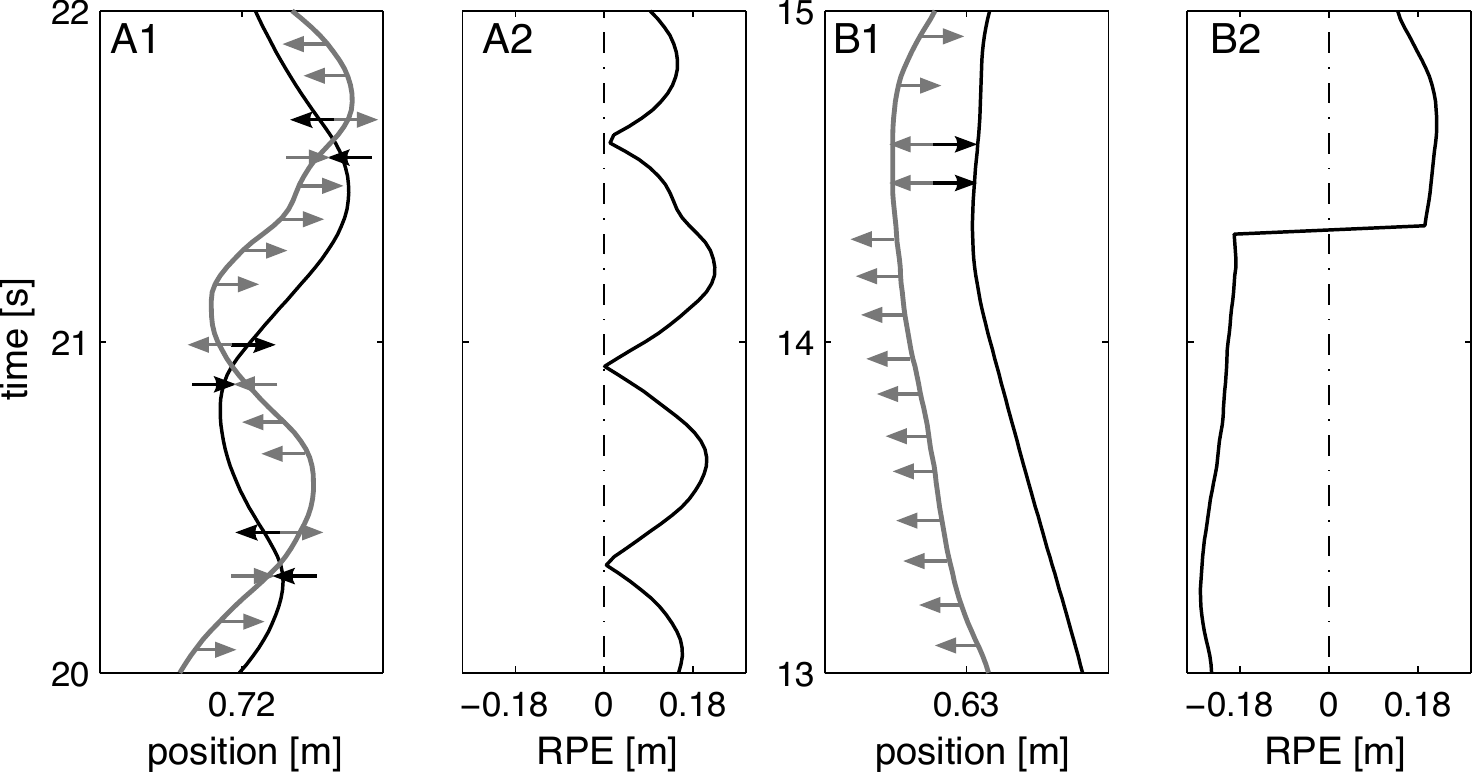}
 \caption{Illustration of the principle used for computing relative position error. In panels A1 and B1 the black lines indicate leader's position and the grey lines indicate follower's position. Grey arrows show direction of motion of the follower. At the times indicated by double grey-black arrows the participants move in opposite directions. Panels A2 and B2 show corresponding relative position error (black). Time runs from bottom to top.}
 \label{fig:rpe_principle}
\end{figure}

The Relative position error (RPE) is based on the notion that if two objects are moving in the same direction then the one behind is following, and on the assumption that changes in direction of movement are initiated by the leader. We define the RPE as the difference in the players' positions multiplied by their common direction of motion. In cases when the players are moving in opposite directions, we assume that the follower is always behind the leader, regardless of the directions of players' movements. These rules lead to the following rule for computing the $\mbox{RPE}({x_1(t),x_2(t)})$:
\begin{equation*}
\begin{cases}
(x_1(t)-x_2(t))\sgn(v_1(t)),&\sgn(v_1(t)) = \sgn(v_2(t))\neq0,\\
|x_1(t)-x_2(t)|,&\mbox{otherwise.}\\
\end{cases}
\label{eq:rpe}
\end{equation*}
Here $x_1, v_1$ are position and velocity of the leader and $x_2, v_2$ are position and velocity of the follower. Positive values of the RPE mean that the follower is behind the leader. We note that the RPE is not symmetric, that is $\mbox{RPE}({x_1(t),x_2(t)}) \neq -\mbox{RPE}({x_2(t),x_1(t)})$ (due to the assumption that a follower should react to the action of a leader) and therefore, it could be treated as a measure of the performance of the follower in addition to indicating the level of temporal correspondence (coordination). 

Figure~\ref{fig:rpe_principle} illustrates the ideas behind computing the RPE. Panels A1 and B1 depict trajectories of movement of two players, leader position is shown with black line, follower position is shown with grey line; time flows from bottom to top. Fig.~\ref{fig:rpe_principle}A2~and~B2 show corresponding time series of the RPE. Grey arrows indicate the direction of motion of the follower. At the times indicated by double grey-black arrows, when the participants move in opposite directions, we compute the RPE using the absolute value of the difference in position, i.e. we assume that the follower is always behind and that the leader initiates changes in the direction of motion.

Figure~\ref{fig:rpe_principle}A shows how our definition of RPE works for the case when the players are often changing direction of motion. In panel A2 we depict a follower that is always behind ($RPE>0$) and that $RPE=0$ occurs at times when the players had the same position but were moving in opposite directions. Fig.~\ref{fig:rpe_principle}B shows that if the follower is ahead of the designated leader the RPE is negative. In such a cases we conclude that the designated leader was tracking the movement of the designated follower. Grey-black arrows in Fig.~\ref{fig:rpe_principle}B1 indicate that in case the designated leader changed direction of movement, the RPE changed sign as well. Comparison of the RPE to other measures of coordination can be found in {\it SI appendix}.

\section{Supplementary Information}

\subsection{Preprocessing}
\begin{itemize}
\item In scenarios 1 and 3, position time series were interpolated with shape-preserving piecewise cubic interpolation (13Hz in experiment 1 and 40Hz in scenario 3).\\ 
Matlab command: \mcode{interp1(t,x,0:1/Fs:t(end),'pchip')}; $Fs$ is the sampling frequency, $t$ is the series of time and $x$ is the position time series.
\item In scenarios 2 and 3, the position data was filtered with a zero-phase forward and reverse digital 2nd order lowpass (10Hz cut-off) Butterworth filter which is a maximally flat magnitude filter. \\ Matlab commands: \mcode{butter(2,10/(Fs/2))} and \mcode{filtfilt}.
\item The pre-processed position time series were used to estimate numerically their corresponding velocity time-series. To differentiate position time-series we used a forth-order finite difference scheme. We cut out the first and last 2 seconds of the signal. Furthermore, we limit velocities to 3.5 [a.u./s] in experiment 1 and to 2.7 [m/s] in scenarios 2 and 3 (higher velocities were considered a results of noise in the collected data).
\item To estimate the PDF of the player's velocity we use normalised histogram of the velocity time series with 101 equally distant bins between -2.7 and 2.7 [m/s] (or -3.5 and 3.5 [a.u./s] in Experiment 1).
\end{itemize}

\begin{suppfigure*}[t!]
 \centering
 \includegraphics[width=\textwidth]{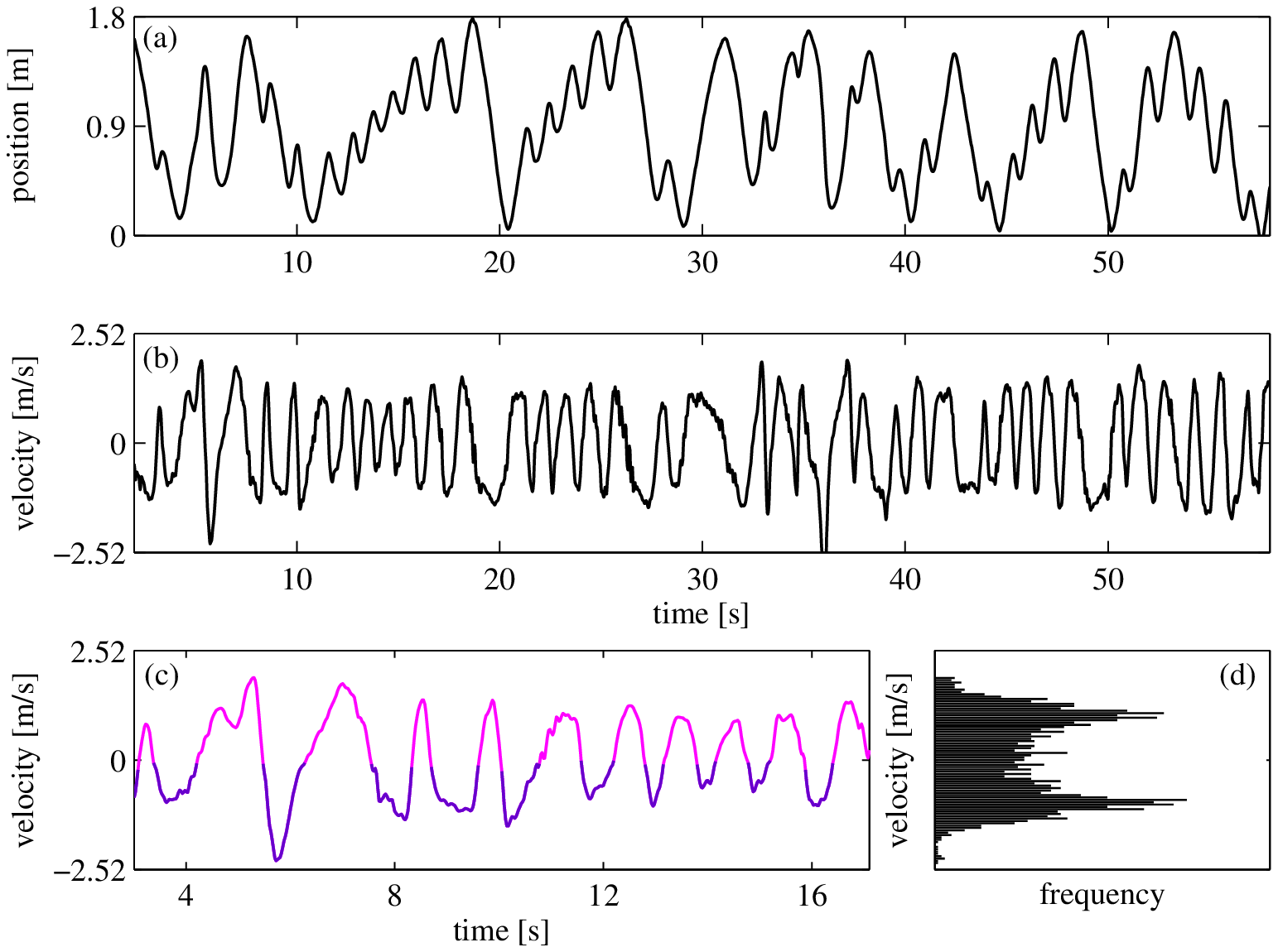}
 \caption{(a) An example of solo position time series from the Experiment 3. (b) Velocity estimated from position data. (c) Fragment of the velocity time series with indicated positive velocity segments (magenta) and negative velocity segments (purple). (d) Velocity profile - histogram of the velocity time series.}
 \label{fig:preprocessing}
\end{suppfigure*}

Supplementary Figure~\ref{fig:preprocessing} illustrates initial stages of analysis of the data.
Supp. Supp.~Fig.~\ref{fig:preprocessing}(a) depicts the representative player's position collected in the solo condition in Scenario 3 and (b) its estimated velocity time-series as explained in  {\it Methods}. For each velocity time-series we compute velocity profile, which is the PDF of the player's velocity time series. In Supp.~Fig.~\ref{fig:preprocessing}(d) we show the velocity profile of the time series represented in Supp.~Fig.~\ref{fig:preprocessing}(b). We use PDFs of velocity in order to capture the essence of the players' movement without being affected by the existing physical constraints on their motion, e.g. limited position range. Multivariate distributions, involving consideration of more than one feature of the motion, would describe the dynamics in a more detailed way but we found that univariate distributions, namely of player's velocities \cite{Hart2014,Hogan1987,Hogan2007,Kilner2003,Kilner2007,Noy2011,Viviani1995} contain enough information to achieve the goals of our study.

Finally, Supp.~Fig.~\ref{fig:preprocessing}(c) depicts the first 20 seconds of the time series from Supp.~Fig.~\ref{fig:preprocessing}(b), with indicated positive velocity segments (magenta) for velocities bigger than 0 that correspond to the movements of the hand from ``left to right'', and negative velocity segments (purple) with velocities smaller than 0 that correspond to movements from ``right to left''. To estimate the PDF of the player's velocity we use normalised histogram of the velocity time series with 101 equally distant bins between -2.7 and 2.7 [m/s] (or -3.5 and 3.5 [a.u./s] in Experiment 1). In Supp.~Fig.~\ref{fig:preprocessing}(d) we show the velocity profile of the time series represented in Supp.~Fig.~\ref{fig:preprocessing}(b).
In order to compare velocity profiles with velocity segments, for each velocity time-series from the experiments 1 and 3 we also find their velocity segments. Velocity segments are fragments of the velocity time series between two consecutive points of zero velocity, i.e. each velocity segments corresponds to a short movement in one direction. For our analysis, we normalise the velocity segments and compute their curve moments. Following \cite{Hart2014,Noy2011} we take into account only velocity segments that are longer than 0.2 sec., shorter than 8 sec. and which have displacement larger than 0.03 [m] (before normalisation). Note that, for the velocity segments, the moments of curve are computed with respect to time and hence parametrise the shape of the velocity segments rather than moments of the sample of velocity (see Section: Moments of a curve of SI).

\subsection{Earth's movers distance}

EMD can be computed using the following Matlab code:
\begin{lstlisting}
bins=linspace(z1,z2,101);	% support Z with 101 bins
bin_width=bins(2)-bins(1);	% widths of the bins, i.e. dz
max_emd=abs(z2-z1); 		% maximal EMD

h1=hist(v1,bins); 	% h1 and h2 are velocity profiles
h2=hist(v2,bins); 	% v1 and v2 are velocity time series
l1=numel(v1); 		% for normalistion
l2=numel(v2); 		% for normalistion
emd_v1v2=sum(abs(cumsum(h1/l1)-cumsum(h2/l2)))*bin_width/max_emd;
\end{lstlisting} 

\subsection{Multidimensional scaling}
We use multidimensional scaling (MDS) to study relations between players' velocity profiles. MDS allows us to model the players' motion as points in an abstract geometric space, which we shall refer to as {\it similarity space}. It is a well established tool in data visualisation and data mining \cite{Borg2005}. It allows to reduce dimensionality of the data and visualise relations between the objects under investigation while preserving as much information as possible. Since the EMD is a metric in the space of velocity profiles (defined by the PDFs of velocity time series), we use classical MDS as implemented in Matlab. We use the Matlab command: \mcode{cmdscale}.

In particular, we first compute the EMDs between all the analysed PDFs, which correspond to individuals' movements. Then we use the computed EMDs in order to construct a matrix {\bf D}. Each row of this matrix is assigned to a different PDF (and hence belongs to a specific individual, i.e. participant in the mirror game) and contains EMDs between this PDF and all the other PDFs. For instance, cell (2,3) contains the EMD between second and third PDFs in our dataset. Since the EMD is a metric, matrix {\bf D} has zeros on the diagonal and is symmetric.

Next, we use the MDS to transform matrix {\bf D} into coordinates of points in the {\it similarity space}. In this way each velocity profile is represented as a single point in the {\it similarity space}. Here we use only the first two dimensions of the {\it similarity space}, which were found to be sufficient for the purpose of our analysis. These two dimensions correspond to the first two highest eigenvalues of matrix {\bf D} computed in the MDS. 

The MDS algorithm is implemented as follows:
\begin{enumerate}
\item Take $n\times n$  matrix $\mathbf{D}$ ($n$ number of analysed objects), and square its elements in order to obtain matrix $\mathbf{D}^{2}$.
\item Transform matrix $\mathbf{D}^{2}$ into matrix $\mathbf{B}$; subtract row means, subtract column means, add back (grand) mean of all the matrix elements and multiply by -0.5. Formally this operation is called double centring and can be expressed as: $\mathbf{B} =-0.5\mathbf{JD}^{2}\mathbf{J}$, here $\mathbf{J} = \mathbf{I} - 1/n \mathsf{11}^T$, where $\mathbf{I}$ is the identity matrix, $\mathsf{1}$ is the vector of ones of length $n$, and $\mathsf{1}^T$ is the transposed vector $\mathsf{1}$.
\item Factor $\mathbf{B}$ by its eigendecomposition $\mathbf{B}=\mathsf{E}\Lambda \mathsf{E}^T$, where $\mathsf{E}$ is matrix which has eigenvectors of $\mathbf{B}$ as columns, and $\Lambda$ is a diagonal matrix with ordered eigenvalues $\lambda_1\geq \lambda_2\geq ... \geq \lambda_n$ on the diagonal.
\item Take the first $m$ eigenvectors $\mathsf{E}_m$ and eigenvalues $\Lambda_m$ of matrix $\mathbf{B}$ and compute $\mathbf{X} = \mathsf{E}_m \Lambda_m^{1/2}$; {\bf X} is a $n\times m$ matrix with $m$ coordinates for each of the $n$ analysed objects; the MDS relies on the property that the eigenvectors of matrix $\mathbf{B}= \mathbf{XX}^T$ can be interpreted as geometric coordinates; $\mathbf{X}^T$ is transposed matrix $\mathbf{X}$.
\end{enumerate}

MDS is a technique related to principal component analysis (PCA) \cite{Borg2005}. In particular, PCA is a statistical procedure which uses singular value decomposition of a matrix $\mathbf{Y}$ or eigendecompostion of covariance matrix $\mathbf{Y}^T\mathbf{Y}/(n-1)$ to study underlying structure of the data. Here $\mathbf{Y}$ is a centred (i.e. its columns have removed means) $n\times m$ matrix of $n$ observation vectors $\tilde{Y}$. In other words, the PCA uses the eigenvectors of the covariance matrix to perform orthonormal transformation of the original coordinate system of the data, i.e. it projects the data into an abstract geometric space with dimensions given by linear combinations of the the original variables. In the same way, the MDS uses eigenvectors and eigenvalues of matrix $\mathbf{B}$ to find a geometric model of the data in an abstract geometric space. The difference between PCA and MDS is the origin and nature of the decomposed matrix. The results of both procedures are eigenvectors and eigenvalues which can be used to reduce the dimensionality of the data while preserving covariance of data (PCA) or preserving distances between analysed objects (MDS). In the case in which distances between analysed objects are given by covariances, i.e. $\mathbf{D}=1-\mathbf{Y}^T\mathbf{Y}$, or if $\mathbf{D}$ is given by euclidian distances between $n$ observations vectors $\tilde{Y}$, both methods give the same results.

The MDS allows us to extend pair-wise analysis of distances between velocity profiles and gain further insight into our data \cite{Slowinski2014}. Furthermore, using MDS guarantees that the euclidian distances between elements in the {\it similarity space} are a good approximation of the EMDs between velocity profiles. Since EMD is a metric we know that the dynamic similarity between players' movements is reflected in the Euclidean distances between their respective  positions in the {\it similarity space}, i.e. the closer the points in {\it similarity space}, the more similar their velocity profiles. This renders the {\it similarity space} a key tool in our analysis of how the dynamic similarity affects mutual rapport and coordination between players in the mirror game.

\subsection{Moments of a curve}
\label{sec:cur_mom}

\begin{supptab}[h]
\caption{Definitions of the first 4 moments of function (curve) $f(t)$ with support/ defined on T=$[{t_1},{t_2}]$.}
\begin{center}
\label{tab:cur_mom}
\begin{tabular}{p{6cm}p{4cm}}
\hline
1st moment --- centre of mass & $\mu=\int\limits_{t_1}^{t_2} t f(t) dt$ \\ \hline
2nd moment --- variance & $\sigma=\int\limits_{t_1}^{t_2} (t-\mu)^2 f(t) dt$ \\ \hline
3rd moment --- skewness & $s=\dfrac{1}{\sigma^{\frac{3}{2}}}\int\limits_{t_1}^{t_2} (t-\mu)^3 f(t) dt$ \\ \hline
4th moment --- kurtosis & $k=\dfrac{1}{\sigma^2}\int\limits_{t_1}^{t_2} (t-\mu)^4 f(t) dt$ \\ \hline
\end{tabular}
\end{center}
\end{supptab}

To analyse and compare movements of different participants Noy~et.~al. use skewness and kurtosis of normalised velocity segments \cite{Hart2014,Noy2011}. To perform a meaningful comparison of the moments of different functions $f(t)$, it is necessary to rescale their supports $T$ to a common one, and to normalise the functions $f(t)$ with their integrals $\int\limits_ {t_1}^{t_2}f(t) dt$. In particular, in \cite{Noy2011} as well as in our analysis the support $T$ of the velocity segments is time normalised into $\tau \in[0,1]$. We note that moments of curve are different from moments of a sample, i. e. they are computed with respect to the support, rather that the values in the sample. For example, in the case of the centre of mass, the area under the curve on the left side of the centre of mass $\mu$ is equal to the area under the curve on the right side of it, that is $\mu$, $\int\limits_{t_1}^\mu f(t) dt=\int\limits_\mu^{t_2} f(t) dt$. 

\begin{suppfigure*}[t!]
 \centering
 \includegraphics[width=\textwidth]{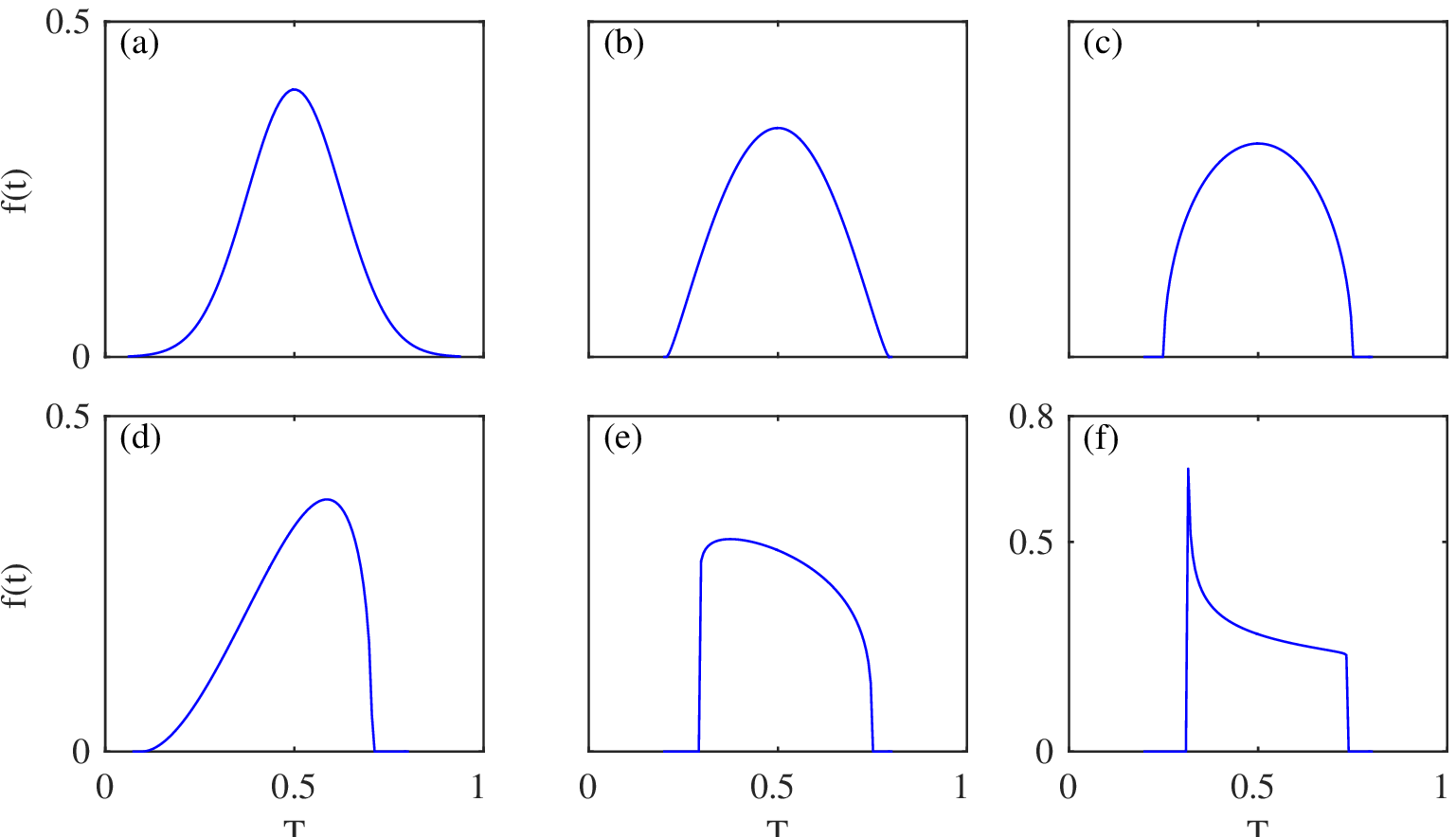}
 \caption{Different curve segments (a) Normal distribution $s=0,k=3$; (b) Minimum jerk $s=0,k=2.2$; (c) $s=0,k=2$; (d) $s=-0.5,k=2.5$; (e) $s=0.15,k=1.9$; (f) $s=0.2,k=1.8$. For all distributions: $\mu=0.5, \sigma=1$.}
 \label{fig:ks_examples}
\end{suppfigure*}

Supplementary Figure~\ref{fig:ks_examples} depicts six examples of normalised curve segments with support $T\in[0,1]$; all presented functions have the same centre of mass $\mu=0.5$ and variance $\sigma=1$, whilst skewness and kurtosis vary between panels. In the case of a velocity segment, skewness indicates asymmetry in acceleration and deceleration, while kurtosis provides information about uniformity of the maximal velocity. Low kurtosis means that an object is quickly accelerating and decelerating and keeps constant velocity in between while high kurtosis means that the object is accelerating slowly, and after reaching maximum velocity it almost immediately starts to slow down; normalised velocity segments with higher kurtosis, generally, have higher maximum velocity.

\subsection{Comparison of velocity segments and velocity profiles}

\begin{suppfigure}[t!]
 \centering
 \includegraphics[width=\textwidth]{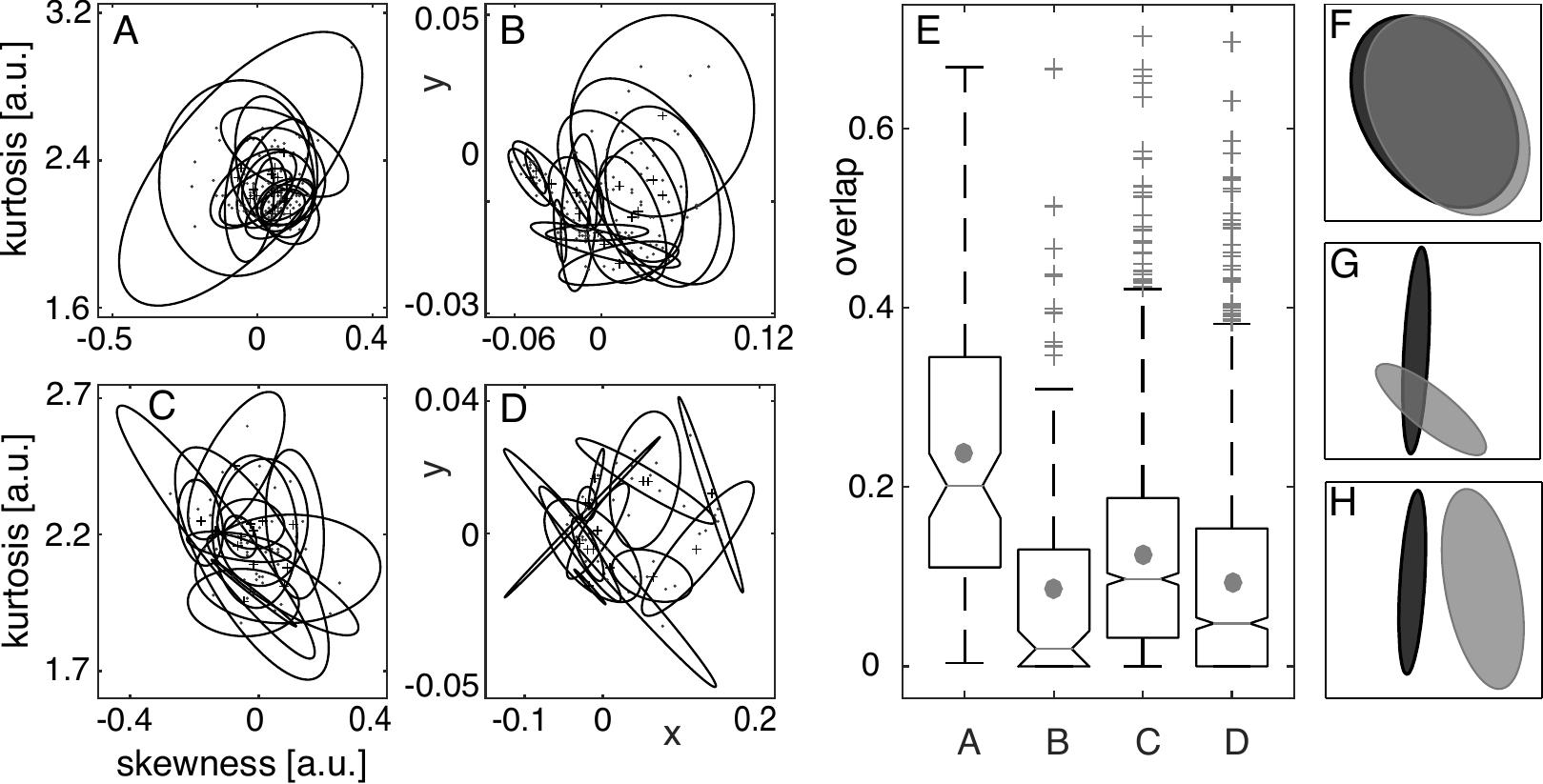}
 \caption{Individual motor signature in A the kurtosis-skewness of velocity segments plane \cite{Hart2014} and in B the {\it similarity space} computed with MDS from distances between velocity profiles. For 15 different participants from experimental Scenario 1, on three different days with at least one week break between recording rounds. Positive velocity segments and velocity profiles from 56 solo trials of 14 participants from experimental Scenario 3 shown in C the skewness-kurtosis plane and in D the {\it similarity space} (for the sake of clarity data for only 14 out of 51 participants is shown). Each ellipse corresponds to a different participant. Small dots correspond to individual solo recordings. Each cross at the centre of an ellipse corresponds to the average of the small dots' positions. Each ellipse indicates 0.7 mass of bivariate normal distribution fitted to the small dots (see {\it SI appendix} for further details). Box plots in panel E show distributions of overlaps $\omega$ between pair of ellipses in panels A--D.  Line between notches indicates median, dot indicates mean. The "central box" represents the central 50\% of the data and its lower and upper boundary lines are at the 25\%/75\% quantile of the data. The two vertical lines extending from the central box indicate the remaining data outside the central box that are not regarded as outliers, crosses indicate outliers. Panels F--G show examples of overlap $\omega$ between pair of ellipses: F the ellipses almost completely overlap, $\omega$=0.91; G the ellipses partially overlap, $\omega$=0.13; H the ellipses do not overlap, $\omega$=0.}
 \label{suppfig:IMS_vel_prof}
\end{suppfigure}

In this section we compare two candidates for the individual motor signature.
Following \cite{Hart2014} we begin our analysis by studying kurtosis and skewness of velocity segments. Velocity segments, which were used to analyse the mirror game in \cite{Hart2014,Noy2011}, are parts of the velocity time series where the participant is moving in one direction, i.e., parts of the velocity time-series between two consecutive times of zero velocity. 
From the viewpoint of motion dynamics, skewness indicates asymmetry in acceleration and deceleration in a velocity segment, while kurtosis provides information about uniformity of the maximal velocity in a velocity segment. Low kurtosis indicates that an object was quickly accelerating and decelerating, and kept maximal velocity for a long time. High kurtosis, on the contrary, means that the object was accelerating slowly, and moved with maximal velocity only for a short period of time.

\subsubsection{Overlap}

In order to analyse separation and clustering of data points corresponding to velocity profiles of individual participants in the {\it similarity space} or on the plane of skewness and kurtosis of velocity segments, we encircle them with ellipses given by bivariate gaussian distribution fitted to their coordinates, and next we compute how much the ellipses overlap. In practice we, first, compute mean values and covariance matrix of coordinates of the $n$ points which we wish to encircle. The points correspond to $n$ trials of a subject. Mean values of the coordinates give the position of the centre of the ellipse, while the eigenvectors of the covariance matrix give directions of major and minor axes of the ellipse. Finally, the lengths of the axes of a covariance ellipse that encloses the desired probability mass are given by the square roots of the eigenvalues of the covariance matrix multiplied by the Mahalanobis radius \cite{Manly2004}. In our analysis we use a radius that encloses all of the data points of a participant, which corresponds to 0.7 of the probability mass of the bivariate normal distribution.

We compute the overlap, $\omega$, between ellipses as a ratio of the area of intersection and the total area of two ellipses. In this way total separation corresponds to no overlap $\omega=0$, whilst complete overlap $\omega=1$ means that we cannot distinguish between the two ellipses and hence we cannot distinguish between points that are encircled by them.  The overlap $\omega$ between ellipses allows to assess clustering and separation between regions of the {\it similarity space}, or the plane of skewness and kurtosis of velocity segments, corresponding to different individuals. The advantage of this simple method is that it is dimension independent and hence allows to compare clustering and overlap in different spaces.

\subsubsection{Comparison}

Supplementary Figure~\ref{suppfig:IMS_vel_prof}B depicts velocity profiles of individual players presented as elements of the {\it similarity space}. Supplementary Figure~\ref{suppfig:IMS_vel_prof}E clearly demonstrates that the median overlap $\tilde{\omega}$ between ellipses, and hence individuals, in the skewness-kurtosis plane $ \tilde{\omega}_{A} = 0.2$ is much higher than the median overlap between ellipses in the {\it similarity space} $\tilde{\omega}_{B} = 0.02$ (significance $p_{AB}<$ 0.0001, Kolmogorov-Smirnov test \cite{Chakravarty1967, Corder2014}). More importantly, there are 45 out of 105 pairs of ellipses that do not overlap at all in the {\it similarity space} while in the kurtosis-skewness plane all pairs of ellipses overlap, ($\min \omega_A=0.004$).
To verify our results we also analyse solo recordings collected in the experimental Scenario 3. In the experimental Scenario 3, players had larger range of movement and all the solo trials of individual players were recorded on a single day. Nevertheless in this case we also find that the median overlap $\tilde{\omega}$ between ellipses in the skewness-kurtosis plane $\tilde{\omega}_{C} = 0.1$ is much higher than the median overlap between ellipses in the {\it similarity space} $\tilde{\omega}_{D} = 0.05$ (significance $p_{CD}<$ 0.0001) and the number of non-overlapping pairs of ellipses is higher in the {\it similarity space} (418 against 183 out of 1378 pairs); see Supp.~Fig.~\ref{suppfig:IMS_vel_prof}C and D. In both experimental scenarios, we observe that separation between ellipses, and hence individuals, is significantly better in the {\it similarity space}. 

Our analysis reveals that, despite being a good source of information about human movement on a short time-scale (rates of acceleration, uniformity of maximal velocity), velocity segments are not specific enough to study the effects of dynamic similarity between individual players. More specifically, we find that although the mean values of kurtosis and skewness of velocity segments exhibit clustering for each person, and that the clusters are preserved over time, there also exists a big overlap between mean skewness and kurtosis of different players, meaning that it is not possible to distinguish between them; see Supp.~Fig.~\ref{suppfig:IMS_vel_prof}A, C and E.

\begin{suppfigure*}
 \centering
 \includegraphics[width=\textwidth]{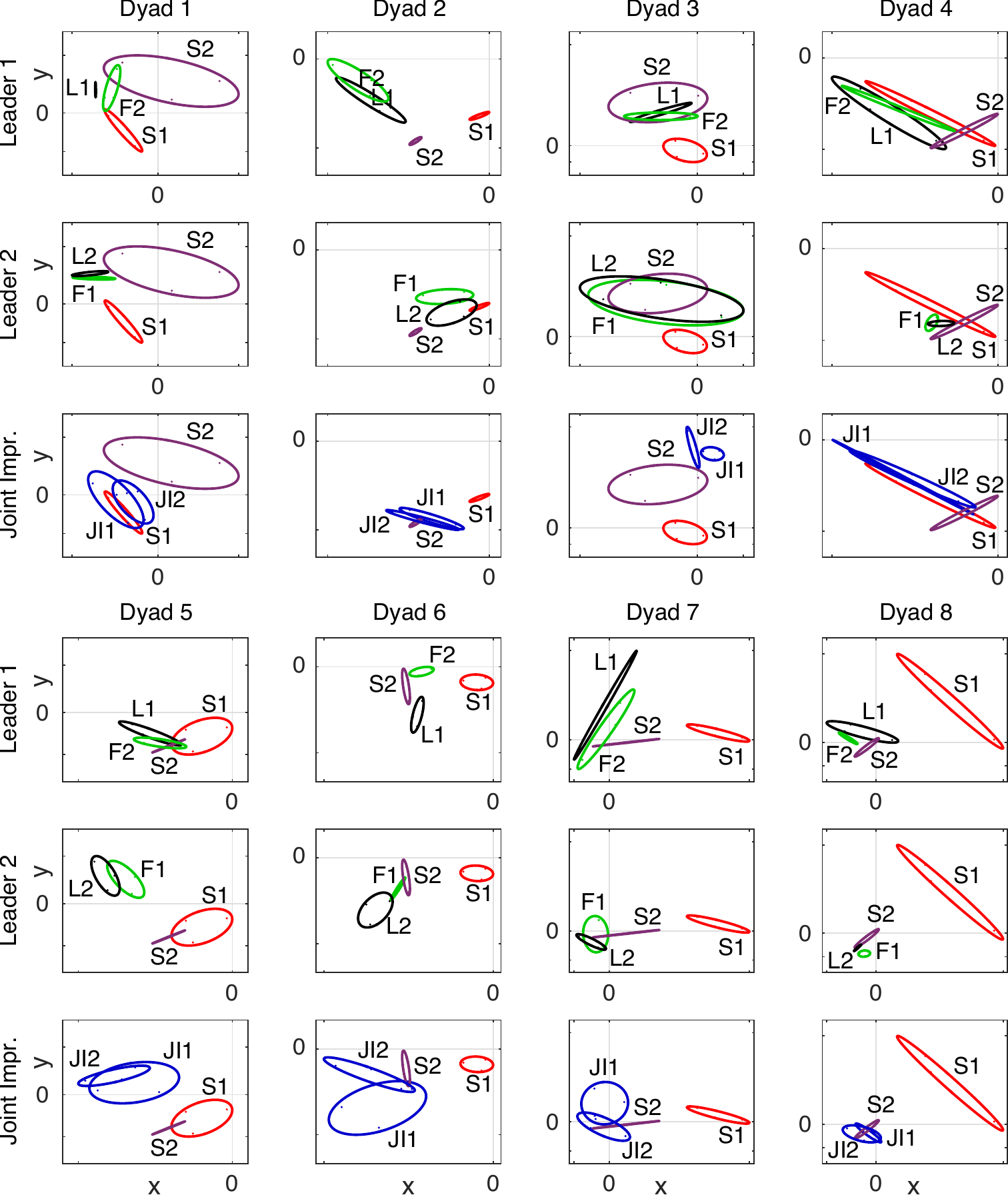}
 \caption{Interaction between two players in different experimental conditions (Scenario 2) visualised in the {\it similarity space} for all 8 dyads. Ellipses encircle points corresponding to velocity profiles in solo (S1 and S2; red), leader (L1 and L2; black), follower (F1 and F2; green) and joint improvisation (JI1 and JI2; blue) rounds. Each column depicts data for two different dyads. $x$ and $y$ axis are rescaled for clarity of presentation.}
 \label{fig:8_dyads}
\end{suppfigure*}

\clearpage

\subsection{Velocity segments of movement generated by ICA}

In this section we use kurtosis and skewness of velocity segments to show that the trajectories generated by the ICA \cite{Zhai2014cdc,Zhai2014smc} in Scenario 3, have the features of a human movement. In our analysis we compared skewness and kurtosis of: velocity segments of human solo movement, velocity segments of human leader movements, and velocity segments of motion generated by the ICA as a leader. We found that except for few outliers, the velocity segments generated by the ICA have the same kurtosis and skewness as the one observed in human motion. 

\begin{suppfigure*}[t]
 \centering
 \includegraphics[width=\textwidth]{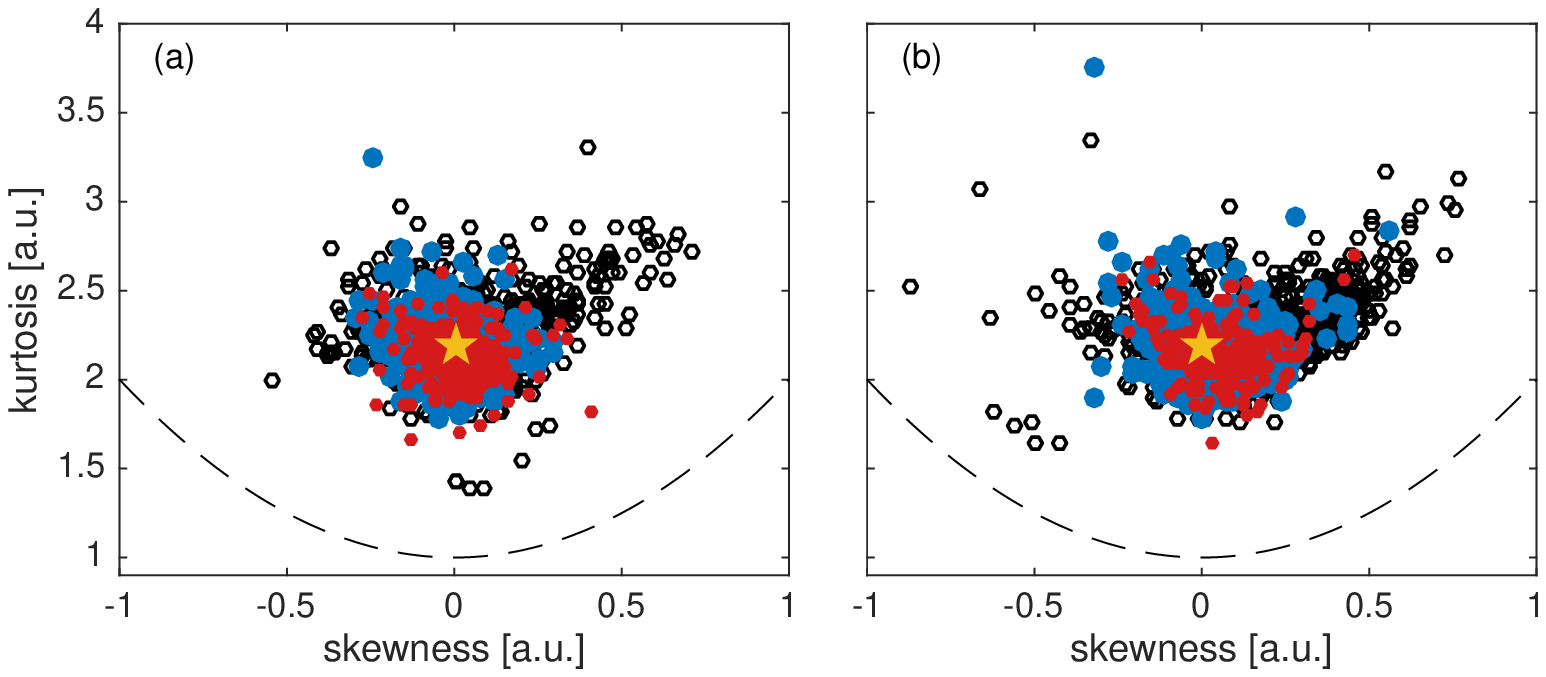}
 \caption{Kurtosis and skewness for velocity segments from time series recorded in the experiment. In panel (a) values from positive velocity segments, in (b) from negative velocity segments; compare with Supp.~Fig.~\ref{fig:preprocessing}(c). In black skewness and kurtosis of velocity segments from ICA acting as a leader. In blue skewness and kurtosis of velocity segments from human players acting as a leader. In red skewness and kurtosis of velocity segments from human players in solo conditions. Yellow star indicates point with $k=2.2$ and $s=0$ which corresponds to the smoothest movement \cite{Hart2014}. Dash-dotted line shows theoretical bound given by relation between kurtosis and skewness of a curve $k \leq s^2 + 1$ \cite{Pearson1929}.}
 \label{fig:LF_Noy}
\end{suppfigure*}

Supplementary Figure~\ref{fig:LF_Noy} shows kurtosis and skewness of velocity segments from time series recorded in experiment 3, (a) for positive velocity segments and (b) for negative velocity segments. Corresponding to the movement of the hand towards and away from the centre of the body, respectively. The skewness and kurtosis of velocity segments from human solo is depicted in red, human leader in blue and avatar leader in black. We notice that for some points the skewness of the avatar leader (black) have bigger values than typical human leader (blue), this is due to the fact that, while leading the human follower, the avatar was at the same time tracking the fast changing pre-recorded reference trajectory. In other words, the stronger skewness indicates that the avatar was quickly accelerating to match the pre-recorded position and then slowly decelerating to allow human follower to catch up. Nevertheless, overall the black dots and blue circles occupy a similar region of the kurtosis-skewness plane. In both panels all the velocity segments are centred on the point with skewness $s = 0$, and kurtosis $k = 2.2$, which is indicated by a cyan star. This point corresponds to the velocity segment with the smoothest movement as reported in \cite{Hart2014,Noy2011}. Normal distribution has skewness $s = 0$, and kurtosis $k = 3$. Dash-dotted line shows theoretical bound of the values of skewness and kurtosis given by the theoretical relation between them $k \leq s^2 + 1$ \cite{Pearson1929}. 

\begin{supptab}[h!]
\caption{Skewness of positive $s_{+}$ and negative $s_{-}$ velocity segments from different movement recordings: S - human solo movement, $L_h$ - human leader, $F_a$ - avatar follower, $L_a$ - avatar leader, $F_h$ - human follower. $p_{ks}$ is significance level of Kolmogorov-Smirnov test.}
\begin{center}
\label{tab:top_bottom}
\begin{tabular}{cp{1.5cm}p{1.5cm}p{1.8cm}}
\hline
& $s_{+}$ &$s_{-}$ & p-value \\ \hline \hline
$S$& 0.0098 & 0.0489 & $p_{ks}$=9.5e-4 \\ \hline
$L_h$& -0.0070 & 0.0291&$p_{ks}<$0.0001 \\ \hline
$L_a$& 0.0473& 0.1320&$p_{ks}<$0.0001 \\ \hline
$F_h$& 0.0913& 0.0658&$p_{ks}$=7.8e-4 \\ \hline
\end{tabular}
\end{center}
\end{supptab}

In Supplementary Table~\ref{tab:top_bottom} we report the difference in skewness of the positive and negative velocity segments. This difference is a result of asymmetry in the movement of the hand towards and away from the centre of the body, i.e., the movement is actuated by different groups of muscles \cite{Flash1985}. We have not found difference between kurtosis of positive and negative velocity segments in any condition.

In summary, Supp.~Fig.~\ref{fig:LF_Noy} shows that in most trials, velocity segments of the avatar leader (driven by ICA) have kurtosis and skewness which are very similar to the human players. Furthermore, higher than in Solo condition values of skewness of the human follower's velocity segments are consistent with the observation that after noticing changes of direction of the leader's movement, the human follower reacts and accelerates quickly to correct her/his position. Next, she/ he slows down to track the leader's movement in a more precise way.

\subsection{Correlations and partial correlations between different variables and RPE}

\begin{supptab}[h]
\caption{Partial correlations between RPE and EMD controlled for $\mu |V_L|$, and RPE and average absolute velocity of the leader $\mu |V_L|$ controlled for EMD in data from scenario 2 and scenario 3.}
\begin{center}
\label{tab:part_corr}
\begin{tabular*}{\hsize}{@{\extracolsep{\fill}}p{2cm}p{4cm}p{4cm}}
Scenario 2:& $\mu |V_{L}|$ & $EMD(S_1,S_2)$\\ \hline \hline
$RPE(L,F)$&  $\rho$=0.3448 ($p_{\rho}$=0.0189)& $\rho$=0.3466 ($p_{\rho}$=0.0183) \\ 
\hline
\\
Scenario 3:& $\mu |V_{LVP}|$ & $EMD(Ref,S)$\\ \hline \hline 
$RPE(L_{VP},F_H)$& $\rho$=0.6554 ($p_{\rho}$=0)& $\rho$=0.1863 ($p_{\rho}$=0.4e-05) \\ 
\hline 
\end{tabular*}
\end{center}
\end{supptab}

Considering the characteristics of motion, the effect of adding a 2.5Hz sinusoidal signal to the solo trajectories results in higher average absolute velocity of the reference trajectory: $R^2(\mu EMD(Ref,S), \mu|V_{Ref}|- \mu|V_S|)$=0.9676 (p = 0), Spearman's $\rho(\mu EMD(Ref,S), \mu|V_{Ref}|- \mu|V_S|)$=0.9689 (p = 0). 
Nevertheless, we still found significant effect of the dynamic similarity controlled for average absolute velocity of the leader, see Supplemetary table~\ref{tab:part_corr}, Scenario 3. The existence of this correlation confirms that the velocity of the leader's motion alone could no explain the variability in the RPE, even in the very limited case when dynamic similarity simplifies to differences between preferred (solo) velocities of the players. 

\begin{supptab}[h]
\caption{Correlations between RPE and different measures of similarity between solo recordings. Relation between data is measured with Pearson $R^2$ and Spearman's rank correlation coefficient $\rho$. $\mu |V_{S\cdot}|$ is average absolute solo velocity, $\max |V_{S\cdot}|$ is maximum absolute solo velocity, $sk_{S\cdot}$ is mean skewness of the solo velocity segments, $kr_{S\cdot}$ is mean kurtosis of the solo velocity segments, $(sk,kr)_{S\cdot}$ are the coordinates of a centre of player's ellipse in the skewness-kurtosis plane obtained from the solo trials.}
\begin{center}
\label{tab:corr_S}
\begin{tabular*}{\hsize}{@{\extracolsep{\fill}}p{4.5cm}p{3cm}p{3cm}}
\hline
& $\mu RPE(L,F)$ & $\mu RPE(L_{VP},F_H)$\\ \hline \hline
$EMD(S_L,S_F)$&  $R^2$=0.3701 ($p_{R^2}$=0.0105)& $R^2$=0.2343 ($p_{R^2}$=4.5e-09)\\ 
&  $\rho$=0.3907 ($p_{\rho}$=0.0066) & $\rho$=0.2224 ($p_{\rho}$=2.7e-08) \\ 
\hline 
$\operatorname{abs}(\mu |V_{SL}|-\mu |V_{SF}|)$&  $R^2$=0.3469 ($p_{R^2}$=0.0169)& $R^2$=0.3281 ($p_{R^2}$=8.0e-17) \\ 
&  $\rho$=0.3453 ($p_{\rho}$=0.0175) &  $\rho$=0.2997 ($p_{\rho}$=3.6e-14)\\ 
\hline 
$\operatorname{abs}(\max |V_{SL}|-\max |V_{SF}|)$&  $R^2$=0.1110 ($p_{R^2}$=0.4576)& $R^2$=0.1154 ($p_{R^2}$=0.0043)\\ 
&  $\rho$=0.1383 ($p_{\rho}$=0.3538)& $\rho$=0.0824 ($p_{\rho}$=0.0415)\\ 
\hline 
$\operatorname{abs}(sk_{SL} - sk_{SF})$&  $R^2$=0.3834 ($p_{R^2}$=0.0078)& $R^2$=0.2124 ($p_{R^2}$=1.1e-07)\\ 
&  $\rho$=0.3152 ($p_{\rho}$=0.0309)& $\rho$=0.1427 ($p_{\rho}$=4.0e-04) \\ 
\hline 
$\operatorname{abs}(kr_{SL} - kr_{SF})$&  $R^2$=0.0686 ($p_{R^2}$=0.6468)& $R^2$=-0.2271 ($p_{R^2}$=1.3e-08) \\ 
&  $\rho$=0.1405 ($p_{\rho}$=0.3461)& $\rho$=-0.2287 ($p_{\rho}$=1.0e-08) \\ 
\hline 
$||(sk,kr)_{SL}- (sk,kr)_{SF}||_2$&  $R^2$=0.1058 ($p_{R^2}$=0.4791)& $R^2$=-0.1346 ($p_{R^2}$=8.4e-04)\\ 
&  $\rho$=0.2132 ($p_{\rho}$=0.1501)& $\rho$=-0.0835 ($p_{\rho}$=0.0389) \\ 
\hline 
\end{tabular*}
\end{center}
\end{supptab}

\clearpage 

\subsection{Statistical tests}

Since our data is not normally distributed, to test the existence of correlations we use Spearman's rank correlation coefficient $\rho$ \cite{Corder2014}. Additionally, for illustrative purposes, we compute the Pearson $R^2$ coefficient of linear dependance. To compute correlation coefficients and their significance values (p-values) we use Matlab commands: \\
\mcode{[R2,p]=corr(x,y,'type','Pearson')} and \\ 
\mcode{[rho,p]=corr(x,y,'type','Spearman')}. \\
For the same reason, to test statistical significance of differences between distributions we use Kolmogorov-Smirnov test computed with Matlab command \mcode{kstest2}. Kolmogorov-Smirnov test determines whether independent random samples are drawn from the same underlying continuous population \cite{Corder2014}.

\subsection{Relative phase}
\label{sec:RP}

Analysis of the relative phase between two (or more) oscillators is an established method for quantifying synchronisation (coordination) level and hence temporal correspondence between periodic time series \cite{Equis2010,Varlet2011}. We performed such analysis by using a method of reconstructing phase of an oscillator from data as described in \cite{Kralemann2008,Revzen2008}. In particular, following \cite{Kralemann2008}, we computed protophase using the Hilbert transform and transformed it into phase, which grows linearly with time, using the Damoco 2 toolbox for Matlab \cite{Damoco2}. However, measures of temporal correspondence that rely on the relative phase based on the Hilbert transform are not suited for the analysis of the time series recorded in our experiments for the following reasons:
\begin{itemize}
\item they have non-zero local mean; signal with moving averages have big jumps in phases which introduce big errors in relative phase (see Fig.~2 in \cite{Equis2010}), 
\item their amplitude and phase spectra are not well separated; relative phase is sensitive to changes of amplitude,
\item in many cases the time series contain multiple frequencies; instantaneous phase based on the Hilbert transform can be computed but does not have a physical interpretation. 
\end{itemize}
More information about issues stated in the above list and importance of different assumptions for correct estimation of the phase of a signal can be found in \cite{Equis2010}. 

Since the results obtained from the analysis of relative phase based on Hilbert transform were not satisfactory, we decided to use a method of estimating the relative phase based on a wavelet transform of a time series \cite{Schmidt2014, Issartel2006,Issartel2015}. In particular, we used estimation of relative phase based on wavelet coherence as described in \cite{Grinsted2004} and implemented in Crosswavelet and Wavelet Coherence toolbox for Matlab \cite{wtc}. Wavelet coherence can be considered a localised correlation coefficient in time-frequency space. 

Wavelet coherence provided us with an estimate of relative phase in the time-frequency space, i.e. at each time we have multiple values of relative phase that correspond to different frequencies. To reduce dimensionality of the time frequency estimate of the relative phase, we averaged it over frequencies, obtaining in this way the time course of relative phase. 

\begin{suppfigure*}[t!]
\centering
 \includegraphics[width=\textwidth]{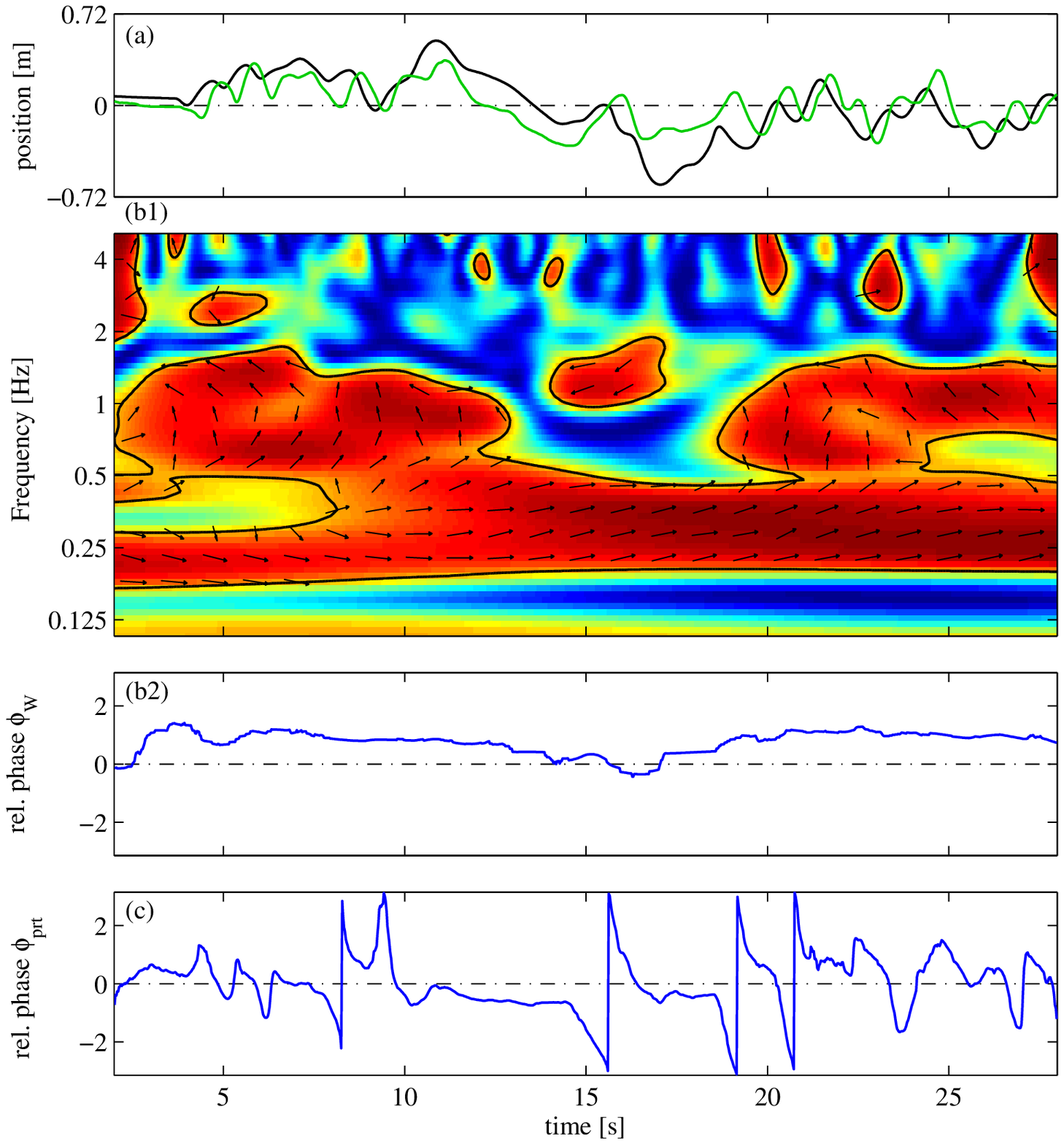}
 \caption{Relation between wavelet coherence and relative phase. (a) Position time series of leader (black) and follower (green). (b1) Squared wavelet coherence between leader and follower time series, red indicates high coherence and blue low coherence. Black contour shows 0.05 significance level, arrows indicate relative phase relationship (clockwise positive values, anti-clockwise negative values). (b2) Relative phase computed with significant wavelet coherence averaged over frequencies. (c) Relative phase between time series based on Hilbert transform.}
 \label{fig:wavelets}
\end{suppfigure*}

Supplementary Figure~\ref{fig:wavelets} illustrates the process of averaging the estimate of the relative phase computed in the time-frequency space over frequency. In particular, Supp.~Fig.~\ref{fig:wavelets}(a) shows the position time series of leader (black) and follower (green), while Supp.~Fig.~\ref{fig:wavelets}(b1) shows the values of the wavelet coherence in the time-frequency space. Red colours indicate high coherence, i.e. the two signals have correlated frequency components at a given time, whilst blue ones indicate regions with no coherence. Black contour delineates the area where common frequencies of both signals are statistically significant; tested against random noise \cite{Grinsted2004}. Arrows are a visualisation of the phase relation between correlated frequency components of the two time series (clockwise angles have negative values, anti-clockwise angles have positive values). Arrows pointing to the right show that the two signals are in-phase. For clarity only arrows in the regions with statistically significant coherence are shown. 

Supp.~Fig.~\ref{fig:wavelets}(b2) shows frequency average of relative phase $\phi_W$ from the regions with statistically significant coherence; we use circular mean to compute the average \cite{Fisher1995}. Supp.~Fig.~\ref{fig:wavelets}(c) shows the relative phase based on the Hilbert transform $\phi_{prt}$ computed with the Damoco 2 toolbox \cite{wtc}; multiple jumps in the relative phase are caused by the changes in the local means of the two signals. This figure clearly demonstrates that the estimate of the relative phase computed with wavelet coherence $\phi_W$ is better than the one based on Hilbert transform $\phi_{prt}$, since the sign of the former is consistent with the fact that the designated leader was actually leading the other player during the joint action. The advantage of this method originates from the fact that $\phi_W$ is based on the parts of the signal which are measurably correlated and can be modelled with periodic functions in the time-frequency plane.

\clearpage

\subsection{Comparison of different measures of temporal correspondence}
\label{sec:diff_tc}

\begin{suppfigure*}[t!]
 \centering
 \includegraphics[width=\textwidth]{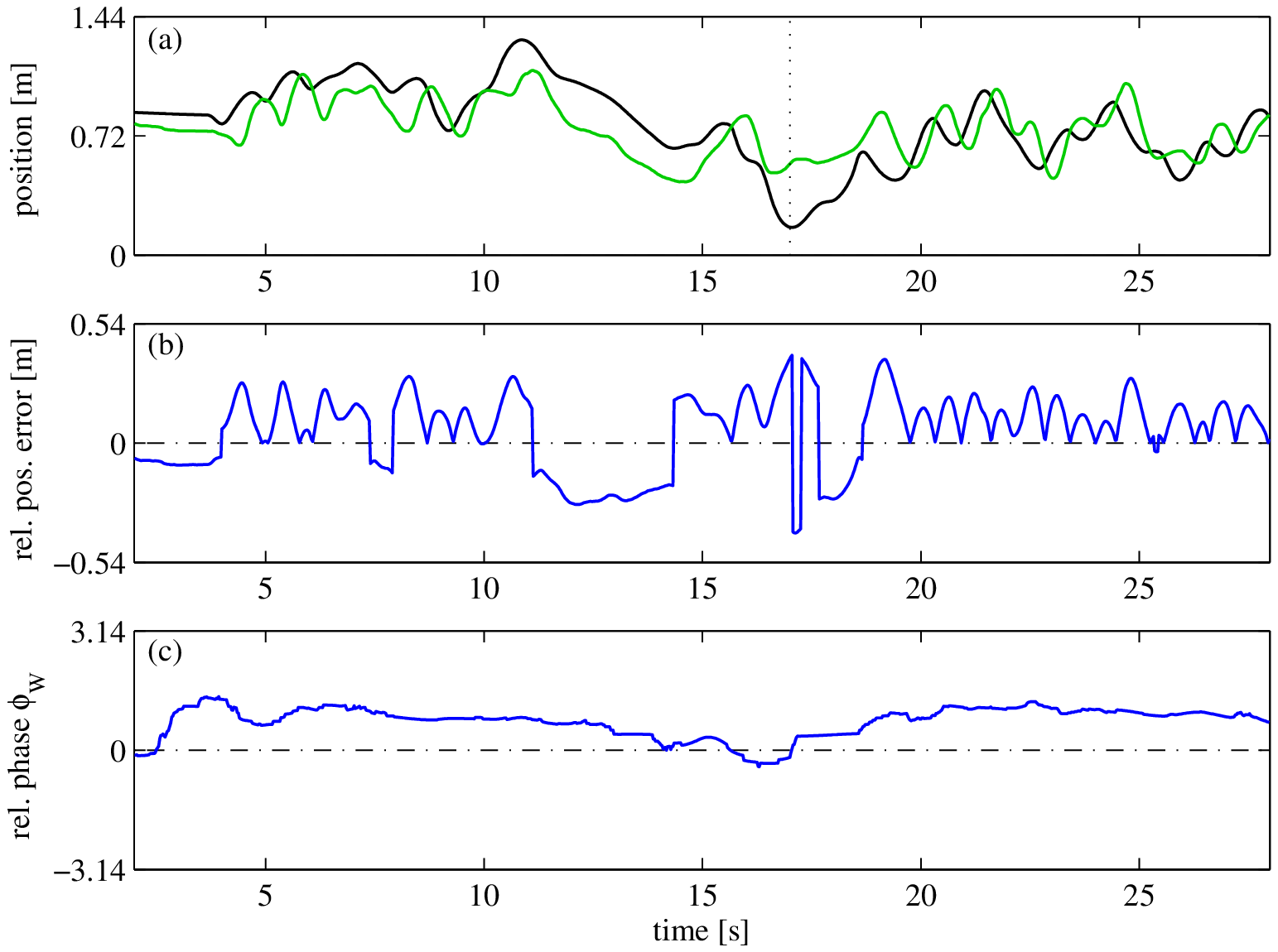}
 \caption{Comparison of relative position error and relative phase computed with wavelet coherence. (a) position time series of leader (black) and follower (green). (b) Relative position error. (c) Relative phase computed with wavelet coherence $\phi_W$.}
 \label{fig:time_lag}
\end{suppfigure*}

Having introduced different measures of temporal correspondence in the sections above 
we now compare the relative error in position and the estimate of the relative phase based on wavelet coherence, using data from experiment 3 (we used data from  experiment 3 because it contains the largest number of trials).

Supplementary Figure~\ref{fig:time_lag}(a) shows position time series of leader (black) and follower (green) (the plots do not start from zero because we cut out the first 2 seconds of the signals). Supp.~Fig.~\ref{fig:time_lag}(b) shows the RPE between the positions in panel (a), while Supp.~Fig.~\ref{fig:time_lag}(c) shows the estimate of relative phase based on the wavelets coherence. We observe in Supp.~Fig.~\ref{fig:time_lag}(b)~and~(c) that the RPE behaves differently compared to the relative phase, e.g. in the time interval [10,15] the RPE indicates that the follower is ahead of the leader ($RPE<0$), while the relative phase shows that there was no exchange of roles between leader and follower. This different behaviour is caused by the fact that the RPE is computed using the information at a given instant of time, while the estimate of the relative phase is based on the wavelet transformation for which time localisation depends on the frequency and is limited by the time-frequency uncertainty principle \cite{Grinsted2004}. Also, in this time interval there are no fast oscillations in the movement, therefore the phase was estimated using low frequency wavelets for which the relative phase was positive (compare with Supp.~Fig.~\ref{fig:wavelets}(b)).

The negative value of the relative phase in Supp.~Fig.~\ref{fig:time_lag}(c) around $t=17s$ is caused by the temporal mismatch between the minimum in the follower's trajectory (green) and the next minimum on the leader's trajectory (black). The minimum in the leader's trajectory indicated by the vertical dotted line in panel (a) occurs after the green one, while all the other extrema of the black trace precede the green trace extrema. The relatively fast change in trajectory in this case was estimated by using higher frequency wavelet of short duration for which the relative phase was negative. Observations from Supp.~Fig.~\ref{fig:time_lag}(c) are consistent with the RPE in panel (b) which has negative values for a short time around $t=17s$.

More generally, based on our analysis and the example discussed above, we conclude that in the context of the mirror game, where the players move along complicated trajectories, the most useful method for quantification and assessment of the leader-follower interaction is the RPE measure. Specifically, the RPE exhibits stronger association to the dynamics of the movement (in terms of statistical significance of the results of the analysis), than the relative phase. Nevertheless, we envisage that the relative phase based on wavelet coherence would be useful when analysing data from the mirror game played in a condition without designated leader, e.g. joint improvisation, when the RPE cannot be used. Finally, an advantage of using the RPE to quantify temporal correspondence is its straightforward physical interpretation. 

\subsection{On the relation between temporal correspondence and dynamic similarity}

\begin{suppfigure}[t!]
\centering
\includegraphics[width=\textwidth]{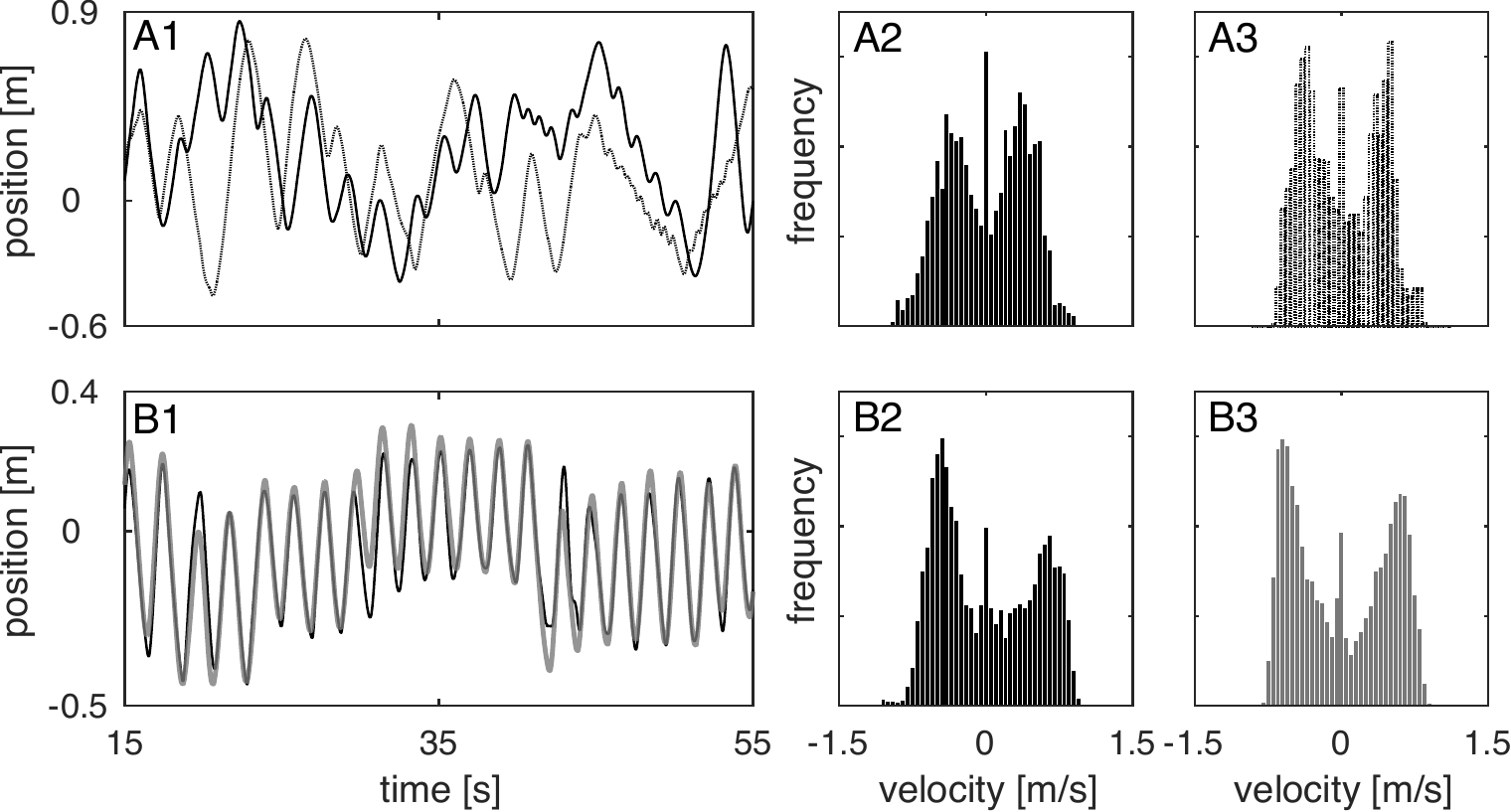}
\caption{Top row shows an example of nontrivial dynamic similarity: panel A1 shows two solo movement trajectories of the same participant, corresponding velocity profiles are shown in panels A2 and A3; $RPE_{A1}$=0.19, $EMD(h_{A2},h_{A3})$=0.009. Bottom row shows an example of trivial dynamic similarity: panel B1 shows movement trajectories of a leader (black) and a follower (grey), corresponding velocity profiles are shown in panels B2 and B3; $RPE_{B1}$=0.008, $EMD(h_{B2},h_{B3})$=0.008.}
 \label{fig:emd_rpe}
\end{suppfigure}

Given two complex time-series, regardless of their origin, it is always possible to measure their temporal correspondence, e.g. using relative position error (RPE), as well as compute their dynamic similarity, using earth's mover distance (EMD) between histograms of their first derivative. By comparing these two quantities, we can define trivial and nontrivial dynamic similarity. 

More specifically, if the two position time series are coordinated, they will necessarily have similar velocity profiles. For example, consider perfect synchronisation when two time series are identical, in such case also their velocity profiles have to be identical and the EMD between them equals 0, i.e. good coordination $\Rightarrow$ small EMD. Dynamic similarity, which is a result of the synchronisation between time series, shall be called trivial. On the other hand, if the EMD between velocity profiles is small, while position time series are uncoordinated we observe nontrivial dynamic similarity i.e. small EMD $\not \Rightarrow$ high coordination. Such situation is possible because velocity profiles do not contain temporal information.

Supplementary Figure~\ref{fig:emd_rpe} illustrates the difference between trivial and nontrivial dynamic similarity. In panel A we depict two trajectories of solo movement of a player and their corresponding velocity profiles. The mean RPE between the two trajectories in panel A1 is equal to $\mu RPE_{A1}=0.19$, and the EMD between histograms in panels A2 and A3 is equal $EMD(h_{A2},h_{A3})=0.009$; it is an example of nontrivial dynamic similarity. In panel B we depict leader (black) and follower (grey) trajectories and their corresponding velocity profiles. Here: $RPE_{B1}=0.008$, and $EMD(h_{B2},h_{B3})=0.008$; panel B shows example of trivial dynamic similarity.

\end{document}